\documentclass[5p,twocolumn,sort&compress,times,pdftex,preprint]{elsarticle}

\usepackage{amsmath, amsfonts, amsthm, amssymb}
\usepackage{color}
\usepackage{ascmac}
\usepackage{url}
\usepackage{ulem}
\usepackage{bm}
\usepackage{bbold}
\usepackage{amsmath}
\usepackage{algpseudocode}

\def\vec#1{\boldsymbol #1}

\newcommand{\la}{\langle}
\newcommand{\ra}{\rangle}

\newcommand{\HPhi}{ \mathcal{H} \Phi}
\newcommand{\Ham}{\hat{\mathcal{H}}}

\newcounter{bla}

\journal{Computer Physics Communications}

\begin{document}

\begin{frontmatter}



\title{Quantum Lattice Model Solver $\HPhi$}


\author[a]{Mitsuaki Kawamura\corref{author}}
\author[a]{Kazuyoshi Yoshimi}
\author[a]{Takahiro Misawa}
\author[b,c,d]{Youhei Yamaji}
\author[e,a]{Synge Todo}
\author[a]{Naoki Kawashima}

\cortext[author] {Corresponding author.\\\textit{E-mail address:} mkawamura@issp.u-tokyo.ac.jp}
\address[a]{The Institute for Solid State Physics, The University of Tokyo, Kashiwa-shi, Chiba, 277-8581,Japan}
\address[b]{Quantum-Phase Electronics Center (QPEC), The University of Tokyo, Bunkyo-ku, Tokyo, 113-8656, Japan}
\address[c]{Department of Applied Physics, The University of Tokyo, Bunkyo-ku, Tokyo, 113-8656, Japan}
\address[d]{JST, PRESTO, Hongo, Bunkyo-ku, Tokyo, 113-8656, Japan}
\address[e]{Department of Physics, The University of Tokyo, Bunkyo-ku, Tokyo, 113-0033, Japan}

\begin{abstract}
$\HPhi$ [\textit{aitch-phi}] is a program package based on the Lanczos-type 
eigenvalue solution applicable to a broad range of quantum lattice models, i.e., 
arbitrary quantum lattice models with two-body interactions, 
including the Heisenberg model, the Kitaev model, the Hubbard model 
and the Kondo-lattice model. While it works well on PCs and PC-clusters, 
$\HPhi$ also runs efficiently on massively parallel computers, 
which considerably extends the tractable range of the system size. 
In addition, unlike most existing packages, $\HPhi$ supports 
finite-temperature calculations through the method of thermal pure quantum (TPQ) states. 
In this paper, we explain theoretical background and user-interface of $\HPhi$.
We also show the benchmark results of $\HPhi$ on supercomputers
such as the K computer at 
RIKEN Advanced Institute for Computational Science (AICS)
and SGI ICE XA (Sekirei) at the Institute for the 
Solid State Physics (ISSP).
\end{abstract}


\end{frontmatter}



{\bf PROGRAM SUMMARY}

\begin{small}
\noindent
{\em Manuscript Title:} Quantum Lattice Model Solver $\HPhi$\\
{\em Authors:}  Mitsuaki Kawamura, Kazuyoshi Yoshimi, Takahiro Misawa, Youhei Yamaji, Synge Todo, Naoki Kawashima \\
{\em Program Title:} $\HPhi$ \\
{\em Journal Reference:}                                      \\
{\em Catalogue identifier:}                                   \\
{\em Program summary URL:} \\
http://ma.cms-initiative.jp/en/application-list/hphi \\
{\em Licensing provisions:} GNU General Public License, version 3 or later\\
{\em Programming language:} C                                   \\
{\em Computer:} Any architecture with suitable compilers including PCs and clusters.\\
{\em Operating system:} Unix, Linux, OS X.  \\
{\em RAM:} Problem dependent. For example, less than one GB for a few-site system, and
3.8 TB for the 36-site Heisenberg model computed in this study.
\\
{\em Number of processors used:} Problem dependent.
 We use up to 4,096 processors (32,768 cores) in this study.  
\\
{\em Classification:}
4.8 Linear Equations and Matrices, 7.3 Electronic Structure
\\
{\em External routines/libraries:} {MPI, BLAS, LAPACK}\\
{\em Nature of problem:}\\
Physical properties (such as the magnetic moment, the specific heat, the spin susceptibility)
of strongly correlated electrons at zero- and finite temperature.
\\
{\em Solution method:}\\
Application software based on the full diagonalization method, the exact diagonalization using the Lanczos method, 
and the microcanonical thermal pure quantum state for quantum lattice model such as the Hubbard model, 
the Heisenberg model and the Kondo model.
\\
{\em Restrictions:}\\
System size less than about 20 sites for a itinerant electronic system and 40 site for a local spin system.
\\
{\em Unusual features:}\\
Finite-temperature calculation of the strongly correlated electronic system
based on the iterative scheme to construct the thermal pure quantum state.
This method is efficient for highly frustrated system which is difficult to
treat with other methods such as the unbiased quantum Monte Carlo.
\\
{\em Running time:}\\
Problem dependent. For example, when we compute the Heisenberg model
on the kagome lattice without $S^z$ conservation,
we can perform 400 iterations per hour.
\\

\end{small}

\section{Introduction}
\label{Intro}
Comparison between experimental observation and theoretical analysis is a crucial step
in condensed matter physics researches. Temperature dependence of the specific heat and
the magnetic susceptibility, for example, has been studied to extract nature of low energy
excitations and magnetic interactions among electrons, respectively, through comparison
with the Landau's Fermi liquid theory \cite{PinesNozieres} and Curie-Weiss law \cite{Kittel}.

For flexible and quantitative comparison with 
experimental data, an exact diagonalization
approach \cite{Dagotto}, which can simulate quantum lattice models
without any approximations 
is one of the most reliable numerical tools.
Most numerical methods for general quantum lattice problems,
which is not accompanied by the negative sign difficulty,
are not exact and reliable error estimate for them is difficult.
This drawback is serious especially when one wants to compare the results 
with experiments. Therefore, exact methods are valuable though 
they are usually available only for small size problems. 
They are important in two ways; as a tool for obtaining 
reliable answers to the original problem (when 
the original problem is a small size problem) 
and as a generator of reference data with which one can 
compare the results of other numerical methods (when the original 
problem is beyond the 
applicable range of the exact methods).

For {the} last few decades, the numerical diagonalization package for quantum
spin Hamiltonians, TITPACK \cite{titpack}
has been widely used in the condensed matter physics community.
Based on TITPACK,
other numerical diagonalization packages
such as KOBEPACK \cite{kobepack} 
and SPINPACK \cite{spinpack} have also been developed.
There is another program performing such a calculation in
the ALPS project (Algorithms and Libraries for Physics Simulations) \cite{1742-5468-2011-05-P05001}.
However, the size of the target system is limited for TITPACK, KOBEPACK, and ALPS
because they 
do not support the distributed memory parallelization.
Although SPINPACK supports such parallelization and an efficient algorithm with the help of symmetries,
this program package cannot handle the 
general interactions 
such as the Dzyaloshinskii-Moriya and the Kitaev interactions that recently
receive extensive attention.
%

In addition,
advances in quantum statistical mechanics \cite{Imada1986,FTLanczos,Hams,Sugiura2012}
open a new avenue for exact methods to finite-temperature calculations
without the ensemble average.
This development enables us 
to calculate finite-temperature properties of 
quantum many-body systems
with computational costs 
similar to calculations of ground state properties.
Now, it is possible
to quantitatively compare theoretical 
predictions for 
temperature dependence of, 
for example, the specific heat and the magnetic susceptibility 
with experimental results \cite{Yamaji2014}.
However, the above program packages have not support this calculation yet.
To perform the large-scale calculations directly relevant to experiments by utilizing 
the parallel computing architectures
with small bandwidth and distributed memory,
a user-friendly, multifunctional, and parallelized diagonalization 
package is highly desirable.

$\HPhi$ [\textit{aitch-phi}], a flexible diagonalization package for solving a
wide range of quantum lattice models,
has been developed
to overcome the problems in the previous program packages.
In $\HPhi$, we implement
the Lanczos method for calculating the ground state 
and properties of a few excited states,
thermal pure quantum (TPQ) states \cite{Sugiura2012}
for finite-temperature calculations,
and full diagonalization method 
for checking results of Lanczos and TPQ methods at small clusters, 
with an easy-to-use and flexible user interface.
By using $\HPhi$, one can analyze a wide range of quantum lattice models including
conventional Hubbard and Heisenberg models, multi-band extensions of the Hubbard model,
exchange couplings that break SU(2) symmetry of quantum spins such as the Dzyaloshinskii-Moriya
and the Kitaev interactions, and the Kondo lattice models describing itinerant electrons coupled to
quantum spins. It is easy to calculate 
a variety of physical quantities such as 
the internal energy, specific heat, magnetization, charge and spin structure factors, and
arbitrary one-body and two-body static Green's functions 
at zero temperature or finite temperatures.

As we mention above,
one of the aims of developing $\HPhi$ is to make a 
flexible diagonalization program package that enables us to directly compare
the theoretical calculations and experimental data.
In addition to the simple model Hamiltonians introduced 
in textbooks of quantum statistical physics,
more complicated Hamiltonians inevitably appear to 
quantitatively describe electronic properties of real compounds. 
In the Lanczos and the TPQ simulation of the quantum lattice model in the condensed matter physics,
the most time-consuming part is the multiplication of the Hamiltonian to a wavefunction.
Excepting the case for the very small system,
it is unrealistic to store all non-zero elements of the Hamiltonian in memory.
Therefore we perform the multiplication by constructing the matrix elements of the Hamiltonian
on the fly.
To carry out the Lanczos and TPQ calculations efficiently,
we implement two algorithms for massively parallel computation; 
one is the conventional 
parallelization based on the butterfly-structured communication pattern
\cite{pacheco1997parallel} (the butterfly method)
and the other is newly developed method
[the continuous-memory-access (CMA) method].
Because the numerical cost of the butterfly method
is proportional to the number of terms in the second quantized Hamiltonian,
it requires very long time to simulate the complicated Hamiltonian 
relevant to the real compounds.
In contrast to the conventional algorithm,
the newly developed method
can realize the continuous memory access and
numerical cost which does not depend on the number of terms
during multiplication between the Hamiltonians and a wavefunction.
In this paper, we also explain these two algorithms
and compare their computational speed.

This paper is organized as follows:
In Sec. 2, we introduce the basic usage of $\HPhi$. 
How to download and build $\HPhi$ are explained in Sec. 2.1,
and how to use $\HPhi$ is explained in Sec. 2.2.
We also explain what types of models can be treated by
using $\HPhi$ in Sec. 2.3.
In Sec. 3, we detail algorithms implemented in $\HPhi$. 
The representation of the Hilbert space adopted in $\HPhi$
is explained in Sec. 3.1.
How to implement
multiplication of the Hamiltonian to a wavefunction
is explained in Sec. 3.2.
Implementation of
full diagonalization method, the Lanczos method, and
the TPQ method is explained in Sec. 3.3, Sec. 3.4., Sec. 3.5,
respectively.
In Sec. 4, we explain the parallelization of
multiplication of the Hamiltonian to a wavefunction.
How to distribute the wavefunction for each process 
is explained in Sec. 4.1.
In Sec. 4.2, we explain the conventional 
butterfly algorithm for
parallelizing the Hamiltonian-wavefunction multiplication.
We also explain
another parallelization method (the CMA method),
which is suitable for treating
complicated quantum lattices models.
In Sec. 5, we show the benchmark results of $\HPhi$.
In Sec. 5.1, we examine the validity of the TPQ method
by using $\HPhi$.
In Sec. 5.2, we show the benchmark results of
parallelization efficiency 
for 18-site Hubbard model and
36-site Heisenberg model on supercomputers.
We also show the benchmark results of 
the CMA method in Sec. 5.3.
Finally, Sec. 6 is devoted to the summary.

\section{Basic usage of $\HPhi$}

\subsection{How to download and build $\HPhi$}

The gzipped tar file,
which contains the source codes, samples, and manuals, can be
downloaded from the $\HPhi$ download site \cite{MA}.
For building $\HPhi$,
a C compiler and the BLAS/LAPACK library\cite{lapack}
are prerequisite. 
To enable the parallel computations,
the Message Passing Interface (MPI) library \cite{MPI} is also required.

For building $\HPhi$, the CMake utility \cite{4052552} can be used as follows:
\begin{verbatim}
$ cd $HOME/build/hphi
$ cmake $PathTohphi
$ make
\end{verbatim}
Here, one 
builds $\HPhi$ in \verb|$HOME/build/hphi| and
it is assumed that the environment variable, \verb|$PathTohphi|,
is set to the path to the source tree path of $\HPhi$ (the top directory of $\HPhi$).
If the CMake utility can not find the MPI library on the system,
the $\HPhi$ executable is automatically compiled without the MPI library.
In this example, CMake will choose a C compiler automatically.
Instead, one can specify the compiler explicitly as follows:
\begin{verbatim}
$ cmake -DCONFIG=$Config $PathTohphi 
$ make
\end{verbatim}
where \verb|$Config| is chosen from the following configurations:
\begin{itemize}
\item \verb|gcc| : GCC
\item \verb|intel| : Intel compiler + MKL library
\item \verb|sekirei| : Intel compiler + MKL library on ISSP system-B (Sekirei)
\item \verb|fujitsu| : Fujitsu compiler + SSL2 library on ISSP system-C (Maki)
\end{itemize}

For a system which does not have the CMake utility, 
we provide another way to generate Makefile's
using \verb|HPhiconfig.sh| shell script. 
One can run \verb|HPhiconfig.sh| in the $\HPhi$ top directory as follows:
\begin{verbatim}
$ bash HPhiconfig.sh gcc
$ make HPhi
\end{verbatim}
Once the compilation finishes successfully, one can find the executable,
\verb|HPhi|, in \verb|src/| subdirectory.

\subsection{How to use $\HPhi$}
Here, we briefly explain how to use $\HPhi$.
$\HPhi$ has two modes; Standard mode and Expert mode.
The difference between them is the format of input files.
For Expert mode, one has to prepare the files in the four categories below.
\\
(1) {\bf Parameter files for specifying the model:} 
In these files, one specifies 
the transfer integrals, 
interactions, 
and types of electronic state (local spins or itinerant electrons).
\\
(2) {\bf Parameter files for specifying the calculation condition:} 
In these files, one specifies the calculation methods (full diagonalization, Lanczos method, and TPQ method), 
the target of models (Spin, Kondo, and Hubbard models), 
number of sites and electrons,
 and the convergence criteria.
\\
(3) {\bf Parameter files for correlation functions:}
By using these input files,
one specifies one-body and two-body equal-time Green's functions to be calculated.
\\
(4) {\bf File for list of above input files:} 
In this file, one should list the names of 
all the files that are necessary
in the calculations. 

For typical models in the condensed matter physics,
such as the Heisenberg model and the Hubbard model,
one can use $\HPhi$ in Standard mode.
In Standard mode,
one can specify all the input parameters
by a single file with a few lines,
from which the files described above are generated automatically
before starting the calculation.
It is thus much more easier to simulate and analyze
these models in Standard mode as long as the model and the lattice is supported in Standard mode.
The models supported in Standard mode will be explained in the following sections.

\subsubsection{Flow of calculations}

A typical flow of calculations in $\HPhi$ is shown as follows:

\begin{enumerate}

\item  Make input files

An example of input file is shown for Standard mode below:
\begin{verbatim}
L       = 4
W       = 4 
model   = "Spin"
method  = "Lanczos"
lattice = "square lattice"
J       = 1.0
2Sz     = 0
\end{verbatim}
Here \verb|L| and \verb|W| are a linear extents of square lattice,
in $x$ and $y$ directions, respectively.
\verb|Spin| means the Heisenberg model, and
\verb|J| is the exchange coupling.
By using this input file, one can obtain the ground state 
of the Heisenberg model for $16=4\times4$ sites 
by performing the Lanczos method.
Details of keywords in the input files can be found 
in the manuals \cite{MA}.

In Expert mode, it is necessary to
prepare all the files that specify the
method, model parameters, and other input parameters.
All the names of input files are listed in
\verb|namelist.def|.
Details of the input files are also found 
in the manuals \cite{MA}.

\item  Run

After preparing the input files,
run a executable \verb|HPhi| in 
terminal by setting option ``\verb|-s|" (or ``\verb|--standard|'') 
for Standard mode and ``\verb|-e|"  for Expert mode as follows:

\begin{itemize}
\item Standard mode

  \verb|$ ./HPhi -s StdFace.def|
\item Expert mode

  \verb|$ ./HPhi -e namelist.def|
\end{itemize}

\item Log and results

Log files and calculation results are output in the 
\verb|output| directory which is 
automatically generated in the working directory. 
For the Lanczos and the full diagonalization methods, 
$\HPhi$ calculates and 
outputs the energy, 
the one-body Green's functions,
and the two-body Green's functions  
from obtained eigenvectors. 
For the TPQ method, the inverse temperature, 
the energy 
and 
its variance 
are also obtained at each TPQ step.
The specified one-body  and two-body Green's
functions are output at specified interval.
\end{enumerate}

\subsection{Models}
Here, we explain what types of models can be treated in $\HPhi$.
In Expert mode, the target models can be generally 
defined by the following Hamiltonian, 
\begin{eqnarray}
  \hat{\cal H} &=&\hat{\cal H}_0 +\hat{\cal H}_{\rm I},
  \\
  \hat{\cal H}_0 &=&\sum_{i j}\sum_{\sigma_1,\sigma_2}
  t_{i \sigma_1 j \sigma_2} \hat{c}_{i\sigma_1}^{\dagger}\hat{c}_{j\sigma_2},
  \\
  \hat{\cal H}_{\rm I} &=&\sum_{i,j,k,l}\sum_{\sigma_1,\sigma_2, \sigma_3, \sigma_4}
  I_{i \sigma_1 j \sigma_2 k \sigma_3 l \sigma_4}
  \hat{c}_{i\sigma_1}^{\dagger}\hat{c}_{j\sigma_2}
  \hat{c}_{k\sigma_3}^{\dagger}\hat{c}_{l\sigma_4},
\end{eqnarray}
where $t_{ij\sigma_1 \sigma_2}$ is a generalized transfer integral 
between site $i$ with spin $\sigma_1$ and site $j$ with 
spin $\sigma_2$ and $I_{ijkl\sigma_1\sigma_2\sigma_3\sigma_4}$ is the 
generalized two-body interaction which
annihilates a spin $\sigma_2$ particle at site $j$ and a spin $\sigma_4$ particle at site $l$, and
creates a spin $\sigma_1$ particle at site $i$ and a spin $\sigma_3$ particle at site $k$.
Here,
$\hat{c}_{i\sigma}^\dag$ ($\hat{c}_{i\sigma}$) is the creation (annihilation) 
operator of an electron on site $i$ with spin $\sigma= \uparrow$ or $\downarrow$.
For the Hubbard and Kondo models, the 1/2-spin fermions are only allowed.
Note here that any system of localized spins can be regarded
as a special case of the above Hamiltonian.
Therefore, one can use $\HPhi$ for solving such quantum spin models
by straight-forward interpretation of the spin interaction in
terms of $t$ and $I$ in the above expressions.
$\HPhi$ also has a capability of handling spins higher than $S=1/2$.

In Standard mode, $\HPhi$ can treat the following models.
(The titles of the following items are the corresponding keys 
for the \verb|model| parameter in a Standard input file.
Here, the keys,
\verb|"Fermion HubbardGC"|, \verb|"SpinGC"| and \verb|"KondoGC"| are
the grand-canonical version of \verb|"Fermion Hubbard"|,
\verb|"Spin"| and \verb|"Kondo"|, respectively.)

\noindent
i) \verb|"Fermion Hubbard"/ "Fermion HubbardGC" |\\
The Hamiltonian is given by 
\begin{align}
\hat{\cal H} &= -\mu \sum_{i \sigma} \hat{c}^\dagger_{i \sigma} \hat{c}_{i \sigma} 
- \sum_{ij, \sigma} t_{i j} \hat{c}^\dagger_{i \sigma} \hat{c}_{j \sigma} \\ \notag
&+ \sum_{i} U \hat{n}_{i \uparrow} \hat{n}_{i \downarrow}
+ \sum_{ij} V_{i j} \hat{n}_{i} \hat{n}_{j},
\end{align}
where $\mu$ is the chemical potential, $t_{ij}$ is the transfer integral 
between site $i$ and site $j$, $U$ is the on-site Coulomb interaction, $V_{ij}$ is the 
long-range Coulomb interaction between $i$ and $j$ sites. 
$\hat{n}_{i\sigma}=\hat{c}_{i\sigma}^\dag \hat{c}_{i\sigma}$ 
is the number operator at site $i$ with spin $\sigma$ and $\hat{n}_i = \hat{n}_{i\uparrow}+\hat{n}_{i\downarrow}$. 
\\ \\
ii) \verb|"Spin"/ "SpinGC"|\\
The Hamiltonian is given by 
\begin{eqnarray}
\hat{\cal H} &= &-h \sum_{i} \hat{S}_{i}^{z} - \Gamma \sum_{i} \hat{S}_{i}^{x} + D \sum_{i} \hat{S}_{i}^{z} \hat{S}_{i}^{z}
\nonumber \\
&+& \sum_{ij, \alpha}J_{i j \alpha} \hat{S}_{i}^{\alpha} \hat{S}_{j}^{ \alpha}+ \sum_{ij, \alpha \neq \beta} J_{i j \alpha \beta} \hat{S}_{i}^{ \alpha} \hat{S}_{j}^{ \beta} ,
\end{eqnarray}
where $h$ is the longitudinal magnetic field, 
$\Gamma$ is the transverse magnetic field, and $D$ is the single-site anisotropy parameter. 
$J_{ij\alpha}$ is the coupling constant between the $\alpha$ component of spins at
site $i$ and site $j$,
and $J_{ij\alpha\beta}$ 
is the coupling constant between the $\alpha$ component of the spin at site $i$
and the $\beta$ component of the spin at site $j$,
where $\{\alpha, \beta\}=\{x, y, z\}$. $\hat{S}_{i}^{\alpha}$ is 
the $\alpha$-axis spin operator at site $i$.
\\\\
iii) \verb| "Kondo"/ "KondoGC"| \\
The Hamiltonian is given by 
\begin{eqnarray}
\hat{\cal H} &=& - \mu \sum_{i \sigma} \hat{c}^\dagger_{i \sigma} \hat{c}_{i \sigma} 
- \sum_{i j,\sigma} t_{ij}\hat{c}^\dagger_{i \sigma}\hat{c}_{j \sigma} 
\nonumber \\
&+& \sum_{i} U \hat{n}_{i \uparrow} \hat{n}_{i \downarrow}
+ \sum_{i j} V_{i j} \hat{n}_{i} \hat{n}_{j}
\nonumber \\
&+& \frac{J}{2} \sum_{i} \left\{
\hat{S}_{i}^{+} \hat{c}_{i \downarrow}^\dagger \hat{c}_{i \uparrow}
+ \hat{S}_{i}^{-} \hat{c}_{i \uparrow}^\dagger \hat{c}_{i \downarrow}
+ \hat{S}_{i}^{z} (\hat{n}_{i \uparrow} - \hat{n}_{i \downarrow})\right\}
\end{eqnarray}
where $\mu$, $t_{ij}$, $U$, and $V_{ij}$ are the same as
those in the ``\verb|Fermion Hubbard|''.
$J$ is the coupling constant between spins of an itinerant electron
and a localized one.

\subsection{Lattice}

\begin{figure}[tb!]
  \begin{center}
    \includegraphics[width=4.0cm]{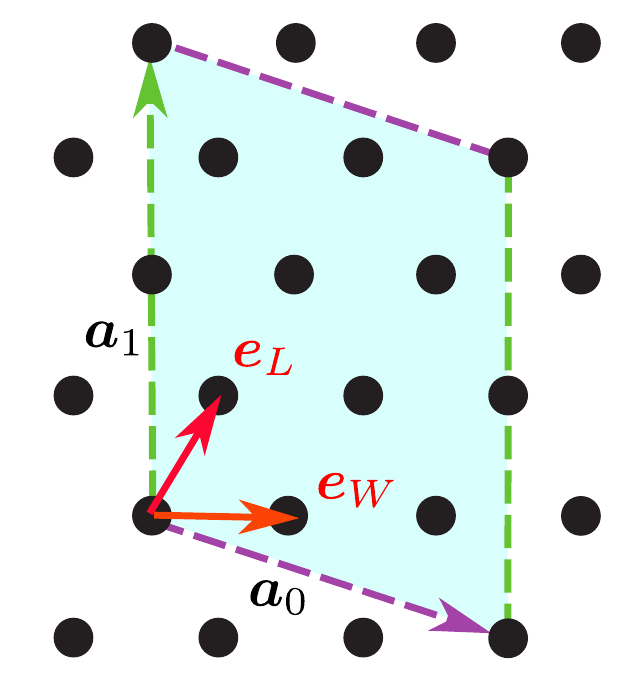}
    \caption{Triangular lattice structure specified by
      \texttt{a0W}=3, \texttt{a0L}=-1,
      \texttt{a1W}=-2, and \texttt{a1L}=4.
      The region spanned by 
      $\vec{a}_0$ (purple dashed arrow) and $\vec{a}_1$ (green dashed arrow)
      becomes the supercell to be calculated (10 sites).
      Red solid arrows ($\vec{e}_W$ and $\vec{e}_L$) are unit lattice vectors.
    }
    \label{fig_triangular}
  \end{center}
\end{figure}

Here, we explain available lattice structures in $\HPhi$.
In Expert mode, arbitrary lattice structures can be 
treated by specifying connections of sites.
In contrast to Expert mode, 
one can easily specify the conventional 
lattice structures
by using Standard mode.
Standard mode supports the chain lattice, the ladder lattice, the square lattice
(examples are shown in insets of Figs. \ref{fig_fdvstpq} and \ref{fig_hubbard_speed}),
the triangular lattice (Fig. \ref{fig_triangular}),
the honeycomb lattice (Fig. \ref{fig_latticegp}), and
the kagome lattice (an example of kagome lattice 
is shown inset of Fig. \ref{fig_kagome}).
In Standard mode, one can construct 
the above two-dimensional lattices 
by specifying four integer parameters, 
\verb|a0W|, \verb|a0L|, \verb|a1W|, and \verb|a1L|
and two unit lattice 
vectors $\vec{e}_W$ and $\vec{e}_L$.
By using these parameters and unit vectors,
a superlattice spanned by
$\vec{a}_0 = a_{0 W} \vec{e}_W + a_{0 L} \vec{e}_L$ and
$\vec{a}_1 = a_{1 W} \vec{e}_W + a_{1 L} \vec{e}_L$
is specified.
We show an example of triangular lattice in Fig. \ref{fig_triangular}.

\subsection{Samples}
There are some sample input files and reference results in \verb|samples/| directory.
For example, 
\verb|samples/Standard/Hubbard/square/| is the directory
containing files for the computation of the Hubbard model on the $2\times4$-site square lattice.
We can perform the calculation of the ground state by running $\HPhi$ as
Standard mode as follows:, 
\begin{verbatim}
$ ./HPhi -s StdFace.def
\end{verbatim}

While running, $\HPhi$ dumps information to the standard output.
We can check the geometry of the calculated system by using \verb|lattice.gp|
which is generated by $\HPhi$; it is read by \verb|gnuplot| \cite{gnuplot} as follows:
\begin{verbatim}
$ gnuplot lattice.gp
\end{verbatim}
\begin{figure}[tb!]
  \begin{center}
    \includegraphics[width=6.0cm]{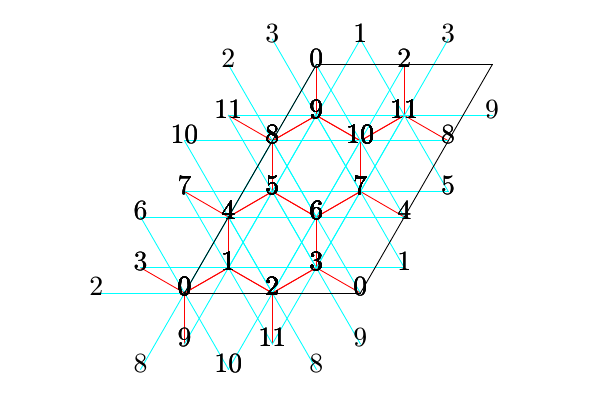}
    \caption{Example of the output of \texttt{lattice.gp} through \texttt{gnuplot}.
      This represents a 12-site honeycomb lattice.
      The corresponding \texttt{lattice.gp} is generated by
      \texttt{samples/Standard/Spin/Kitaev/StdFace.def}.
    }
    \label{fig_latticegp}
  \end{center}
\end{figure}
Then, \verb|gnuplot| displays the shape of the system and indices of sites
(See Fig. \ref{fig_latticegp} as an example.
This figure is obtained by using \verb|samples/Standard/Spin/Kitaev/StdFace.def|).

The calculated results are written in \verb|output/|.
In \verb|output/zvo_energy.dat| the total energy,
the number of doublons (sites occupied by two electrons with opposite spins),
and the magnetization along the $z$ axis are output.
We can compare it with the reference data in \verb|output_Lanczos/zvo_energy.dat| below
\verb|samples/Standard/Spin/Kitaev/|.
Correlation functions are obtained in \verb|output/zvo_cisajs.dat| and
\verb|output/zvo_cisajscktalt.dat|.

When executed in Standard mode, $\HPhi$ creates various input files, 
such as calcmod.def and modpara.def, 
that can be used in Expert mode. 
Editing those files may be the easiest way of using $\HPhi$ for models not supported in Standard mode. 
To do so, after making all the necessary modifications, run $\HPhi$ by the following command:
\begin{verbatim}
$ ./HPhi -e namelist.def
\end{verbatim}
By using Expert mode, one can perform more flexible calculations.

\section{Algorithms implemented in $\HPhi$}
In $\HPhi$, we implement the following three methods;
full diagonalization, 
exact diagonalization 
by the Lanczos method 
for ground state calculation,
the TPQ method for calculation of physical properties at finite temperatures.
{In this section, }
we {first} explain the internal representation of the Hilbert space in $\HPhi$ and
how to implement the multiplication of the Hamiltonian to a wavefunction 
in $\HPhi$.
Next, we explain above three methods.


\subsection{Internal representation of Hilbert space in $\HPhi$}

To specify a state of a site in $\HPhi$, 
we use an $n$-ary, which takes one of $n$ different values. 
For example, in the case of an itinerant electronic system, 
the local site may take four states, $0, \uparrow, \downarrow$, and $\uparrow\downarrow$, 
and therefore it is natural to represent it by a 4-ary. In this representation, 
a basis $|0, \uparrow, \downarrow, \uparrow \downarrow\rangle$
is labeled by
$[0,1,2,3]_{4}{=[0123]_4}=
0\cdot4^3+1\cdot4^2+2\cdot4^1+3\cdot4^0$
as $|0, \uparrow, \downarrow, \uparrow \downarrow\rangle = |[0123]_4 \rangle$.
Here, $[a_{m-1},\cdots,a_1,a_0]_{n}=[a_{m-1}\cdots a_1 a_0]_n$ $(0 \leq a_j < n, j\in [0,m))$
represents the {$m$-digit $n$-ary} number, i.e.,
\begin{align}
  [a_{m-1},\cdots,a_1,a_0]_{n}=[a_{m-1}\cdots a_1 a_0]_n=
  \sum_{j=0}^{m-1} a_{j}\cdot n^{j}.
\end{align}
To use the standard bitwise operations (\textit{e.g.} logical disjunction and exclusive or),
we decompose the 4-ary representations 
as {the following example:}
{A basis $|0, \uparrow, \downarrow, \uparrow \downarrow\rangle$ of the 4-site system is indexed by a quartet of $2$-digit binary numbers as}
\begin{align}
&
{\left[[00]_2,[01]_2,[10]_2,[11]_2\right]_{4}} \notag \\
&=
{[\overbrace{0,0}^{0},\overbrace{0,1}^{\uparrow},\overbrace{1,0}^{\downarrow},\overbrace{1,1}^{\uparrow\downarrow}]_{2}}
\notag \\
&=0\cdot2^{7}+0\cdot2^6+0\cdot2^5+1\cdot2^4 \notag \\
&+1\cdot2^3+0\cdot2^2+1\cdot2^1+1\cdot2^0.   \notag
\end{align}
For the Kondo system, 
the local sites are classified into itinerant electron part and localized spin part.
{However, for simplicity, we use 4-ary representations likewise for the itinerant electronic system, i.e.,}
we {represent}
{a localized spin by}
{$|\uparrow\rangle=|[[10]_2]_{4}\rangle$} {or}
{$|\downarrow\rangle=|[[01]_2]_{4}\rangle$}. 

For localized spin-$S$ systems ($S=1/2,1,3/2 \dots$), 
to represent Hilbert spaces we use $(2S+1)$-ary representation.
For example, in a spin-1 system
the wavefunction is represented as
{$|1,0,-1\rangle=|[2,1,0]_{3}\rangle$}.
To treat the spin-$S$ system, we used Bogoliubov representation
shown in \ref{Bogo}.
In $\HPhi$, we can treat the more complicated
spin systems such as the spin angular momentums have different
values at each site (mixed spin systems). 

In general, at $i$-th site with the spin angular momentum $S_i$, 
the Hilbert space at $i$-th site is represented by $(2S_i+1)$-ary number and 
the Hilbert space of the system is represented 
by $|\phi_{N_{\rm s}-1}, \cdots \phi_1, \phi_0\rangle$
where $\phi_i = \{-S_i, -S_i +1 , \cdots, S_i \}$ and $N_{\rm s}$ is the whole number of sites.
By using the multiplications of the $n$-ary number,
we represent the mixed spin systems as follows:
\begin{equation}
  |\phi_{N_{\rm s}-1}, \cdots \phi_1, \phi_0\rangle
  =\sum_{i=0}^{N_{\rm s}-1} (\phi_i + S_i)\prod_{j=0}^{i-1} (2S_j +1).
\end{equation}

We note that, when the canonical system
is selected by the input file,  $\HPhi$ automatically 
constructs the restricted Hilbert space $\{ \bm{\phi} \}$ 
that has the specified particle number or total $S_z$.
To perform the efficient reverse lookup 
in the restricted Hilbert space,
we use the two-dimensional search method \cite{titpack,Lin}.
We also use the algorithm quoted in \cite{hacker} as
``finding the next higher number after a given number that has the same number of 1-bits''
to perform an efficient two-dimensional search method.

\subsection{Full Diagonalization method}

We generate the matrix of ${\hat{\cal H}}$ by using above-mentioned basis set
$| \psi_j \rangle$
($j=1\cdots d_{\rm H}$, $d_{\rm H}$ is the dimension of the Hilbert space): 
${\cal H}_{ij}= \langle \psi_i | {\hat {\cal H}} | \psi_j \rangle$.
By diagonalizing this matrix,
we can obtain all the eigenvalues $E_{i}$ and
normalized eigenvectors $|\Phi_i\rangle$ ($i=1 \cdots d_{\rm H}$). 
In the diagonalization, we use lapack routine such as \verb|dsyev| or \verb|zheev|.
We also calculate and output
the expectation values $\la A_i\ra \equiv \langle \Phi_i | {\hat A} | \Phi_i\rangle$.
These values are used for the finite-temperature calculations.

From
$\la A_i\ra \equiv \langle \Phi_i | {\hat A} | \Phi_i\rangle$,
we calculate finite-temperature properties by using the ensemble average 
\begin{equation}
\langle {\hat A}\rangle=\frac{\sum_{i=1}^N \la A_i\ra {\rm  e}^{-\beta E_i}}{\sum_{i=1}^N{\rm  e}^{-\beta E_i}}.
\end{equation}
In the actual calculation,
the ensemble averages are performed as a postprocess.

\subsection{Multiplying Hamiltonian to wavefunctions (matrix-vector product)}
\label{sec:multiply}

Here, we explain the main operation in $\HPhi$, multiplying
the Hamiltonian to the wavefunctions.
Because the dimension of the wavefunction becomes exponentially large
by increasing the number of sites, 
it is impossible to store the whole matrix in the actual calculations.
Thus, in the Lanczos (Sec. 3.4) and the TPQ (Sec. 3.5) methods,
we only store the wavefunctions
and calculate the matrix elements at each interaction on the fly.

We explain the procedure of multiplying the Hamiltonian 
to the wavefunctions by taking the 4-site Heisenberg model with $S=1/2$ on the chain 
with periodic boundary condition as an example. 
The Hamiltonian is block-diagonalized with blocks being characterized by the total magnetization,
${\hat S}^z_{\rm Total} \equiv \sum_i {\hat S}^z_i$.
For the block satisfying $\langle {\hat S}^z_{\rm Total} \rangle = 0$,
the wavefunction can be represented by
using the simultaneous eigenvectors of all ${\hat S}^z_i$'s as
\begin{align}
|\phi\rangle
&=a_{3}|\downarrow \downarrow \uparrow\uparrow\rangle
+a_{5}|\downarrow \uparrow \downarrow\uparrow\rangle 
+a_{6}|\downarrow \uparrow \uparrow\downarrow\rangle  \notag \\
&+a_{9}|\uparrow \downarrow \downarrow\uparrow\rangle 
+a_{10}|\uparrow \downarrow \uparrow\downarrow\rangle
+a_{12}|\uparrow \uparrow \downarrow\downarrow\rangle.
\end{align}
Here, the quartet of up or down arrows, which we call a ``configuration'' in what follows,
specifies a basis vector.
In the actual calculations, we store the list of the configurations.

The Hamiltonian is defined as
\begin{align}
  \hat{\cal H}&= \sum_{i=0}^{3}\hat{\vec{S}}_{i}\cdot\hat{\vec{S}}_{{\rm mod}(i+1,4)}
  = \hat{\cal H}_{z}+\hat{\cal H}_{xy},
\end{align}
where $\hat{\cal H}_{z}$ and $\hat{\cal H}_{xy}$ are defined as follows,
\begin{align}
  \hat{\cal H}_{z}  &=  \sum_{i=0}^{3}\hat{S}^{z}_{i}\hat{S}^{z}_{{\rm mod}(i+1,4)},
  \notag \\      
  \hat{\cal H}_{xy} &=  \sum_{i=0}^{3}\frac{1}{2}(
  \hat{S}^{+}_{i}\hat{S}^{-}_{{\rm mod}(i+1,4)}
  +\hat{S}^{-}_{i}\hat{S}^{+}_{{\rm mod}(i+1,4)}
  ).
\end{align}
To reduce the numerical cost, we store
diagonal elements such as $\hat{\cal H}_{z}$ for each configuration.
In contrast to the diagonal terms,
$\hat{S}^{+}_{i}\hat{S}^{-}_{i+1}+\hat{S}^{-}_{i}\hat{S}^{+}_{i+1}$ 
changes the configuration,
e.g. ,
\begin{align}
  (\hat{S}^{+}_{0}\hat{S}^{-}_{1}+\hat{S}^{-}_{0}\hat{S}^{+}_{1})
  |\uparrow\downarrow\uparrow\downarrow\rangle
  =|\uparrow\downarrow\downarrow\uparrow\rangle.
\end{align}
In $\HPhi$, if the exchange operation is possible,
the operation is performed by bit operations as follows:
\begin{align}
[0011]_{2}
\text{\^~}
[1010]_{2}
=[1001]_{2},
\end{align}
where \^~ represents the ``bitwise exclusive or".
Before performing the exchange operator, we count bits in the configuration 
and judge whether the exchange operation is possible or not. If the exchange operation 
is not possible, we do not perform the operation and regard  the contribution from the
configuration as zero, i.e. ,
\begin{align}
  (\hat{S}^{+}_{0}\hat{S}^{-}_{1}+\hat{S}^{-}_{0}\hat{S}^{+}_{1})
  |\downarrow\downarrow\uparrow\uparrow\rangle=0.
\end{align}

We note that the sign 
arises from the exchange relation 
between fermion operators in the 
itinerant electrons systems
such as Hubbard models, while the sign
does not appear in the localized spin systems.
To 
obtain the sign, 
it is necessary to count the parity of
the given bit sequences.
We use the algorithm for parity counting that is
explained in the literature \cite{hacker}.

\subsection{Lanczos method}

For the sake of completeness, we briefly review the principle of the Lanczos method.
For more details see, for example, the TITPACK manual [2] and the textbook by
M. Sugihara and K. Murota \cite{SugiharaMurota_en}.


The Lanczos method is
based on the power method.
In the power method,
by 
successive operations of the Hamiltonian $\Ham$ to the
arbitrary vector $\vec{x}_{0}$, we generate bases $\Ham^{n}\vec{x}_{0}$.
The obtained linear space 
$\mathcal{K}_{n+1}(\Ham,\vec{x}_{0})=\{\vec{x}_{0},\Ham^{1}\vec{x}_{0},\dots,\Ham^{n}\vec{x}_{0}\}$
is called the Krylov subspace.
The initial vector is represented by the superposition 
of the eigenvectors 
$\vec{e}_{i}$ (corresponding eigenvalues are $E_{i}$) of $\Ham$ as 
\begin{align}
\vec{x}_{0}=\sum_{i}a_{i}\vec{e}_{i}.
\end{align}
We note that all the eigenvalues are real number because Hamiltonian is a Hermitian operator.
By operating $\Ham^{n}$ to the initial vector,
we obtain the relation as
\begin{align}
  \Ham^{n}\vec{x}_{0}=E_{0}^{n}
  \Big[ a_{0}\vec{e}_{0}+\sum_{i\neq0}\left(\frac{E_{i}}{E_{0}}\right)^na_{i}\vec{e}_{i}\Big],
\end{align}
where we assume that
$E_{0}$ has the maximum absolute value of the eigenvalues.
This relation shows that
the eigenvector of $E_{0}$ becomes 
dominant for sufficiently large $n$. 
We note that the Krylov subspace does not change under the
constant shift of the Hamiltonian \cite{Frommer}, i.e.,
\begin{align}
\mathcal{K}_{n+1}(\Ham,\vec{x}_{0})=\mathcal{K}_{n+1}(\Ham+aI,\vec{x}_{0}),
\end{align}
where $I$ is the identity matrix and
$a$ is a constant.
By taking constant $a$ as the large positive value,
the leading part of 
$(\Ham+aI)^{n}\vec{x}_{0}$ 
becomes the maximum eigenvectors and vice versa.
Thus, by constructing the Krylov subspace,  
we can obtain eigenvalues around 
both the maximum and the minimum eigenvalues.

In the Lanczos method,
we successively generate the normalized orthogonal basis 
${\vec{v}_{0},\dots,\vec{v}_{n-1}}$ from the Krylov subspace $\mathcal{K}_{n}(\Ham,\vec{x}_{0})$.
We define the initial vector and associated components as 
$\vec{v}_{0} =\vec{x}_{0}/|\vec{x}_{0}|$,
$\beta_{0}=0,\vec{x}_{-1}=0$.
From this initial condition,
we can obtain the normalized orthogonal basis as follows:
\begin{align}
\alpha_{k} &= (\Ham\vec{v}_{k},\vec{v}_{k}), \\
\vec{w}   &= \Ham\vec{v}_{k}-\beta_{k}\vec{v}_{k-1}-\alpha_{k}\vec{v}_{k}, \\
\beta_{k+1} &= |\vec{w}|, \\
\vec{v}_{k+1} &= \frac{\vec{w}}{|\vec{w}|}.
\end{align}
From these definitions, it is obvious that $\alpha_{k}$ and $\beta_{k}$ are real number.

In the subspace spanned by these normalized orthogonal basis,
the Hamiltonian is transformed as
\begin{align}
T_{n}=V_{n}^{\dagger}\Ham V_{n}.
\end{align}
Here,
$V_{n}$ is the matrix whose column vectors are 
$\vec{v}_{i}\quad(i=0,1,\dots,n-1)$.
We note that $V^{\dagger}_{n}V_{n}=I_{n}$ ($I_{n}$ is an $n\times n$ identity matrix).
$T_{n}$ is a tridiagonal matrix and its diagonal elements
are $\alpha_{i}$ and
subdiagonal elements are $\beta_{i}$.
It is known that
the eigenvalues of $\Ham$ are well approximated by 
the eigenvalues of $T_{n}$ for sufficiently large $n$.
The original eigenvectors of $\Ham$ is obtained 
by $\vec{e}_{i}=V_{n}\tilde{\vec{e}}_{i}$,
where $\tilde{\vec{e}}_{i}$ are
the eigenvectors of $T_{n}$.
From $V_{n}$, 
we can obtain the eigenvectors of $\Ham$.
However, in the actual calculations,
it is difficult to keep $V_{n}$ because its dimension
is {too} large to store [dimension of $V_{n}$ = (dimension of the total Hilbert space) 
$\times$ (number of Lanczos iterations)].
Thus, to obtain the eigenvectors, 
in the first Lanczos calculation, 
we keep $\tilde{\vec{e}}_{i}$
because its dimension is small (upper bound of the dimensions 
of $\tilde{\vec{e}}_{i}$ is the number of Lanczos iterations).
Then, we again perform the same Lanczos calculations
after we obtain the eigenvalues from the Lanczos methods.
From this procedure, we obtain the eigenvectors from $V_{n}$.

In the Lanczos method, by successive operations of the Hamiltonian on the initial vector,
we obtain accurate eigenvalues around
the maximum and minimum eigenvalues and associated eigenvectors
by using only two vectors whose dimensions are the dimension of the total Hilbert space.
In $\HPhi$, to reduce the numerical cost,
we store two additional vectors; a vector for accumulating the 
diagonal elements of the Hamiltonian, 
and another vector for the list of the restricted Hilbert space
explained in Sec. \ref{sec:multiply}. 
The dimension of these vectors is that of the Hilbert space.
As detailed below,
to obtain the eigenvector by the Lanczos method,
one additional vector is necessary.
Within some number of iterations,
we obtain accurate eigenvalues near 
the maximum and minimum eigenvalues.
The necessary number of iterations is small enough 
compared to the dimensions
of the total Hilbert space.
We note that it is shown that
the errors of the maximum and minimum eigenvalues
become exponentially small as a function of Lanczos iterations
(for details, see Ref. \cite{SugiharaMurota_en}).
Details of generating the initial vectors and
convergence conditions are shown in 
\ref{ape:IniLz} and \ref{ape:ConvergenceLz}, respectively.

\subsubsection{Inverse iteration method}

In some cases, the accuracy of 
obtained eigenvectors by the Lanczos method 
is not enough for calculating the correlation functions. 
To improve the accuracy of the
eigenvectors, we implement the inverse iteration 
method in $\HPhi$, which is also implemented in 
TITPACK \cite{titpack}.

By using the approximate value 
of the eigenvalues $E_{n}$,
by successively operating $(\Ham-E_{n})^{-1}$
to the initial vector $\vec{y}_{0}$,
we can obtain the accurate eigenvector for $E_{n}$.
The linear simultaneous equation for such procedure is given by
\begin{align}
\vec{y}_{k}&=(\Ham-E_{n})\vec{y}_{k+1}.
\end{align}
By solving this equation by using the
conjugate gradient (CG) method, 
we can obtain the eigenvector.
We note that we take $\vec{y}_{0}$ as the eigenvector
obtained by the Lanczos method and add small positive 
number to $E_{n}$ for stabilizing the CG method.
Although we can obtain more accurate eigenvector by using this method,
additional four vectors are necessary to
perform the CG method.

\subsection{Finite-temperature calculations by TPQ method}

The method of the TPQ states is based on the fact \cite{Hams}
that it is possible to calculate the finite-temperature properties from a few wavefunctions 
(in the thermodynamic limit, only one wavefunction is necessary) 
without the ensemble average.
In what follows we describe the prescription proposed in \cite{Sugiura2012}, 
which we adopt in $\HPhi$.
Such wavefunctions that replace the ensemble average are called TPQ states.
Because a TPQ state can be generated by operating the Hamiltonian 
to the random initial wavefunction,
we directly use the routine of the Lanczos method to the TPQ calculations.
Here, we explain how to construct the TPQ state,
which offers a simple way for 
finite-temperature calculations.

Let $|\psi_{0}\rangle$ be a random initial vector.
How to generate $|\psi_{0}\rangle$ is described in \ref{ape:IniTPQ}.
By operating $(l-\hat{\cal H}/N_{s})^{k}$ ($l$ is a constant,
$N_{s}$ represents the number of sites) 
to $|\psi_{0}\rangle$,
we obtain the $k$-th TPQ states as
\begin{align}
|\psi_{k}\rangle \equiv \frac{(l-\hat{\cal H}/N_{s})|\psi_{k-1}\rangle}{|(l-\hat{\cal H}/N_{s})|\psi_{k-1}\rangle|}.
\end{align}
From  $|\psi_{k}\rangle$, we estimate the corresponding inverse temperature $\beta_{k}$ as
\begin{align}
\beta_{k}\sim \frac{2k/N_{s}}{l-u_{k}},~~
u_{k} = \langle \psi_{k}|\hat{\cal H}|\psi_{k}\rangle/N_{s},
\end{align}
where $u_{k}$ is the internal energy.
Arbitrary local physical properties at $\beta_{k}$ are also estimated as 
\begin{align}
\langle \hat{A}\rangle_{\beta_{k}} =  \langle \psi_{k}|\hat{A}|\psi_{k}\rangle/N_{s}.
\end{align}

In a finite-size system,
a finite statistical fluctuation
is caused by the choice of the initial random vector.
To estimate the average value and the error of the physical properties,
we perform some independent calculations 
by changing $|\psi_{0}\rangle$.
Usually, we regard the 
standard deviations of the physical properties
as the error bars.

\section{Parallelization}

\subsection{Distribution of wavefunction} \label{sec_dist_wave}
In the calculation of the exact diagonalization based on the Lanczos method or
the finite-temperature calculations based on the TPQ state,
we have to store the many-body wavefunction 
on the random access memory (RAM).
This wavefunction becomes exponentially 
large with increasing the number of sites.
For example, when we compute 
the 40-site $1/2$ spin system in the $S_z$ unconserved 
condition, its dimension is 
$2^{40}=1,099,511,627,776$ 
and its size in RAM
becomes about 17.6 TB (double precision complex number).
Such a large wavefunction can be treated by distributing it to many processes by utilizing distributed memory parallelization.

\begin{figure}[tb!]
  \includegraphics[width=9.0cm]{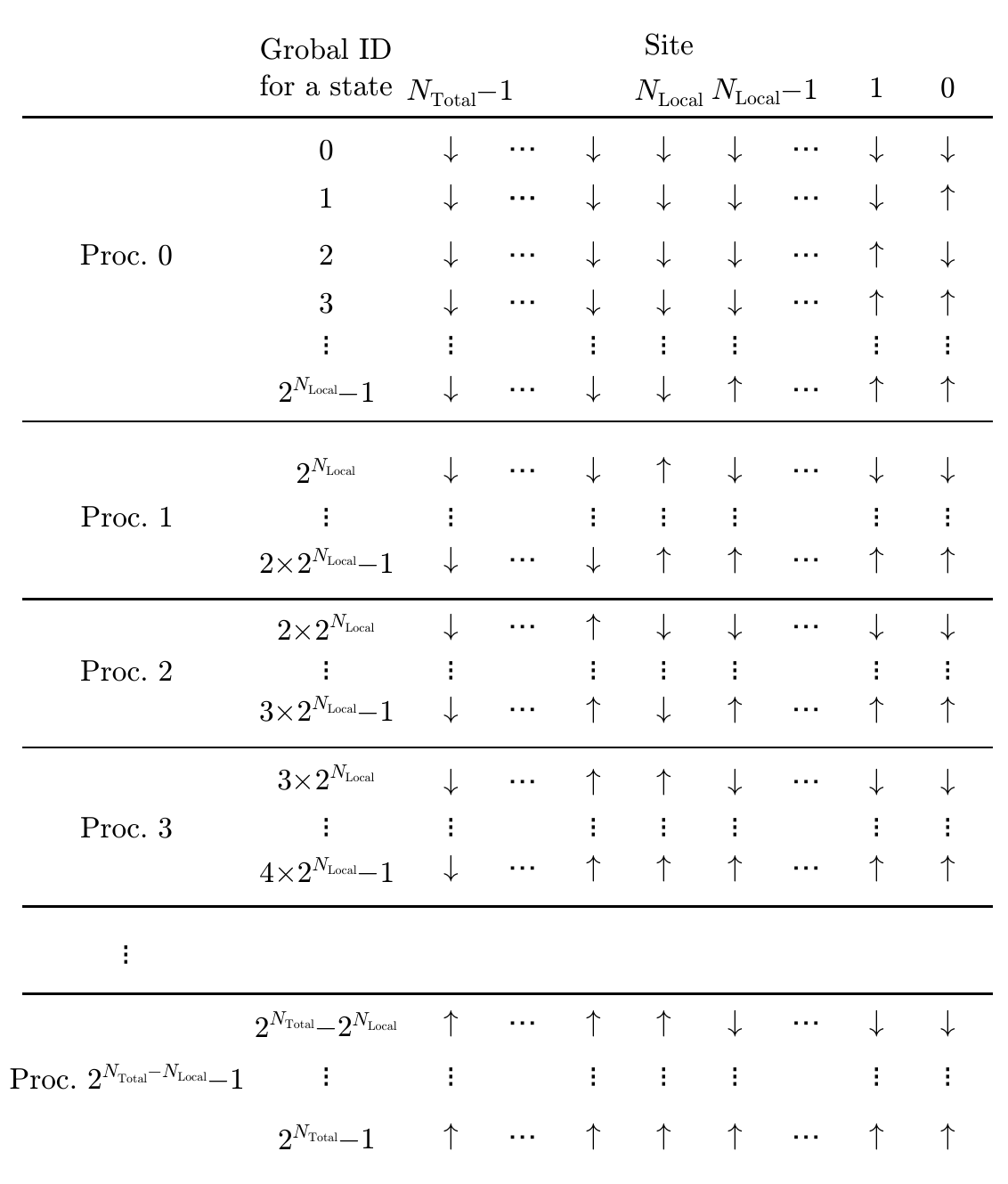}
  \caption{Distribution of the wavefunction of
    the $N_{\rm Total}$-site $1/2$ spin system in the $S_z$ unconserved condition.
    Each process has the wavefunction having $2^{N_{\rm Local}}$ component;
    we can obtain the wavefunction of the entire system by
    connecting local wavefunction in all processes.
  }
  \label{fig_parallel}
\end{figure}

In $\HPhi$, {the ranks of the processes are labeled by using the leftmost bits in
the bit representation of the wavefunction basis.
When the leftmost bits corresponding to a $N_{\rm p}$-site subsystem are chosen to label the ranks of the processes,
the number of processes have to be 
$(2S+1)^{N_{\rm p}}$ (for spin systems) or $4^{N_{\rm p}}$ (for itinerant electron systems).
Then, each process keeps partial wavefunction
of the complementary $N_{\rm Local}(\equiv N_{\rm s}-N_{\rm p})$-site subsystem.
Distribution of wavefunctions of 
an $N_{\rm total} (\equiv N_s)$
spin-1/2 system is illustrated in Fig. \ref{fig_parallel}.}
We note that the configuration of the 
$N_p$ sites is fixed
in the same process.

For example, we show the distribution of the wavefunction of
the $N_{\rm Total}$-site $1/2$ spin system in the $S_z$ unconserved condition 
in Fig. \ref{fig_parallel} (the number of processes is $2^{N_{\rm Total} - N_{\rm Local}}$).
Each process has the wavefunction having $2^{N_{\rm Local}}$ components and
we can obtain the wavefunction of the entire system by
connecting local wavefunctions of all processes.

\subsection{Parallelization of Hamiltonian-vector product}
\label{ParaHvprod}
\begin{figure*}[tb!]
  \includegraphics[width=18.0cm]{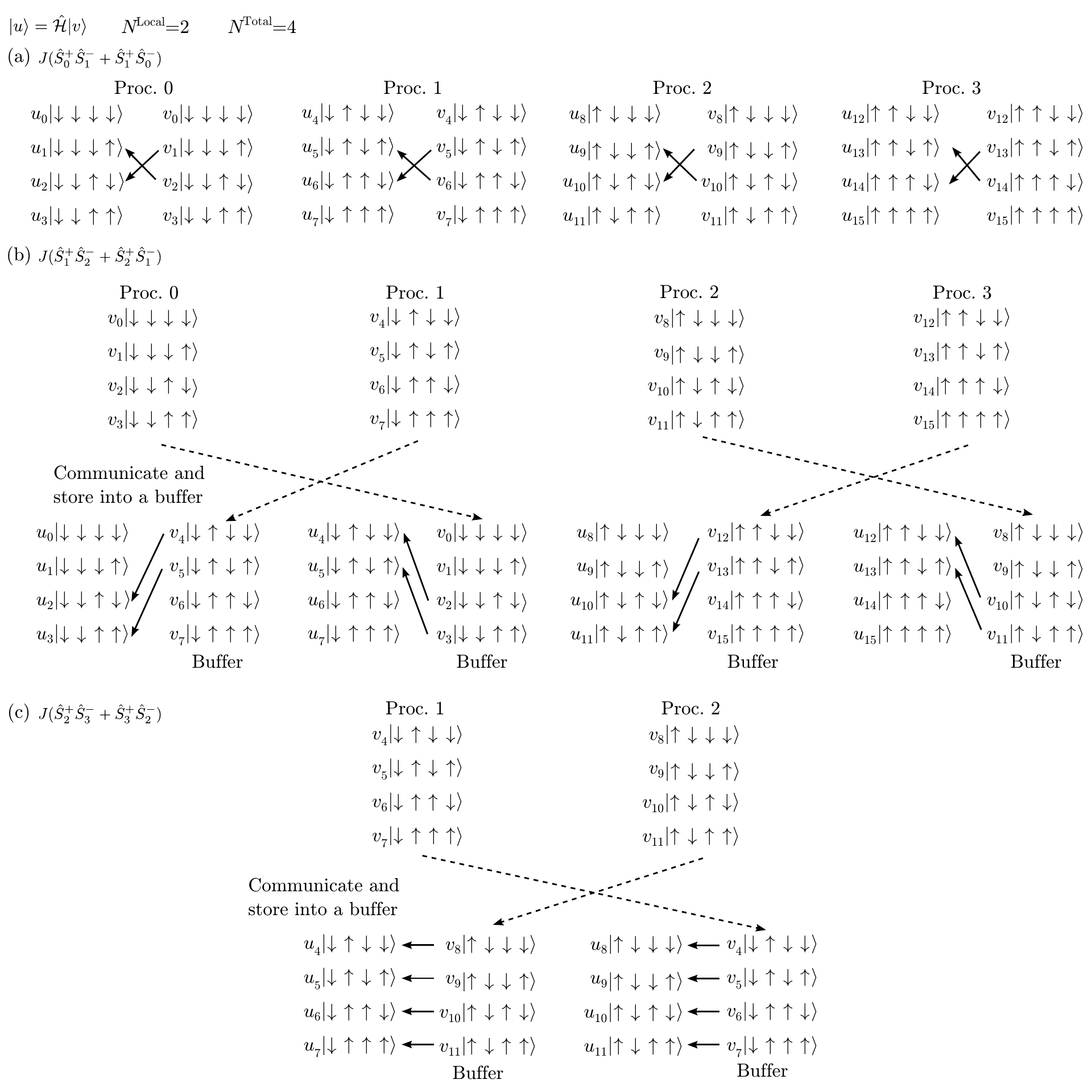}
  \caption{Schematic picture of parallelization of Hamiltonian-wavefunctions product.
  We take a 4-site spin-1/2 Heisenberg model as an example.}
  \label{fig_commun}
\end{figure*}

The most time-consuming procedure in a calculation of the Lanczos method
or the TPQ state is the multiplication of the Hamiltonian with a vector
($|u\rangle = {\hat {\mathcal H}} |v\rangle$),
where $|u\rangle\equiv\sum_{n:{\rm config.}} u_n |n\rangle$ and $|v\rangle$ are wavefunctions distributed to all the
processes.
We parallelize this procedure with the aid of 
Message Passing Interface (MPI) \cite{MPI}.

Here, we explain the parallelization procedure by taking
a spin exchange term
as an example.
Because the spin exchange occurs between two sites,
we use one of the following three procedures according to the
type of the two sites (See Fig. \ref{fig_commun}).
When 
the numbers specifying the both sites
are smaller than 
$N_{\rm Local}$,
we can perform this exchange independently in each process;
there is no inter-process communication [Fig. \ref{fig_commun}(a)].
If one of two sites has an index {equal to or} larger than $N_{\rm Local}$,
first the local wavefunction is communicated between two processes connected
with the exchange operator and stored in a buffer.
Then we compute the remaining exchange operator [Fig. \ref{fig_commun}(b)].
We employ this procedure also when both sites 
are {equal to or} larger than $N_{\rm Local}$ [Fig. \ref{fig_commun}(c)].
In this case, some processes do not participate the calculation;
it causes a load imbalance. 
However, it is not so serious problem;
the computational time of this case is shorter than
those time of the previous two cases
because the arrays are accessed sequentially in this case.

\subsection{Continuous-memory-access (CMA) method}
\label{CA}

Numerical costs become smaller if all operations can be done similar to Fig. \ref{fig_commun}(a)
in which the memory access is limited within each process.
In addition, if we can apply simultaneously the all terms in the second quantized Hamiltonian 
acting to the part of wavefunctions continuously stored in the CPU-cache memory,
the numerical cost becomes small and independent from the number of the terms.
Indeed, we can partially convert the conventional algorithm into an efficient algorithm
by employing a permutation of the bits.
By the permutation, we can also achieve continuous memory access,
which is usually faster than random access
happening inevitably in the conventional algorithm (Sec. \ref{ParaHvprod}).
In this section, we explain the newly developed algorithm to realize
the continuous memory access
and to enhance the computing efficiency.

Any partial Hamiltonians acting on the rightmost $M$ bits
are represented by a $2^M$ by $2^M$ matrix that acts on $2^M$ successive components,
namely, $0$ th to $(2^M-1)$-th components, of the wavefunction.
Then, if the $W$ ($\leq M$) rightmost bits are permuted in the bit representation of the basis as,
\begin{eqnarray}
  &&|[\sigma_{N_{\rm s}-1}\cdots\sigma_{W+1}\sigma_{W}\sigma_{W-1}\cdots\sigma_{1}\sigma_{0}]_2\rangle
  \nonumber\\
  &\rightarrow&
  |[\sigma_{W-1}\cdots\sigma_{1}\sigma_{0}\sigma_{N_{\rm s}-1}\cdots\sigma_{W+1}\sigma_{W}]_2\rangle,
\end{eqnarray}
where $\sigma_i =0,1$ ($i\in [0,N_{\rm s})$),
the partial Hamiltonians acting on the $W$ th bit, $(W+1)$ th bit, and so on,
are again represented by a matrix that acts on the successive components of the wavefunction. 
If we find an integer $W$ that satisfies ${\rm mod}(N_{\rm s},W)=0$
and decompose the total Hamiltonian into the partial Hamiltonians
that act on sets of $M$ successive bits that overlap each other,
the multiplication of the Hamiltonian is implemented with continuous memory access
by repeating the $W$-bits permutation.

When the wavefunctions are distributed in many processes,
the $W$-bits permutation
is achieved by the following steps.
First, we permute the components in the wavefunction
within the rank $\ell$ 
process,
$v^{(\ell)
}_j$
($j=0,1,\cdots,2^{N_{\rm s}-N_{\rm p}}-1$), with a buffer
$u^{(\ell)
}_j$
as
\begin{eqnarray}
\begin{array}{ll}
&
u_{
[
{\sigma}_{W-1}
\cdots
{\sigma}_{W-N_{\rm p}}
{\sigma}_{W-N_{\rm p}-1}
\cdots
{\sigma}_{0}
{\sigma}_{N_{\rm s}-N_{\rm p}-1}
\cdots
{\sigma}_{W}]_2}^{
([
{\sigma}_{N_{\rm s}-1}
\cdots
{\sigma}_{N_{\rm s}-N_{\rm p}}
]_2)}
\\
&=
v_{
[
{\sigma}_{N_{\rm s}-N_{\rm p}-1}
\cdots
{\sigma}_{W}
{\sigma}_{W-1}
\cdots
{\sigma}_{W-N_{\rm p}}
{\sigma}_{W-N_{\rm p}-1}
\cdots
{\sigma}_{0}
]_2}^{
([
{\sigma}_{N_{\rm s}-1}
\cdots
{\sigma}_{N_{\rm s}-N_{\rm p}}
]_2)}.
\end{array}
\end{eqnarray}
{Then, we call an MPI all-to-all routine to transfer the components in the buffer $u^{(\ell)}_{j}$
to another buffer ${u'}^{(\ell')}_{j'}$ in other processes as}
\begin{eqnarray}
\begin{array}{ll}
&
{u'}_{
[
{\sigma}_{N_{\rm s}-1}
\cdots
{\sigma}_{N_{\rm s}-N_{\rm p}}
{\sigma}_{W-N_{\rm p}-1}
\cdots
{\sigma}_{0}
{\sigma}_{N_{\rm s}-N_{\rm p}-1}
\cdots
{\sigma}_{W}]_2}^{
([
{\sigma}_{W-1}
\cdots
{\sigma}_{W-N_{\rm p}}]_2)}
\\
&=
u_{
[
{\sigma}_{W-1}
\cdots
{\sigma}_{W-N_{\rm p}}
{\sigma}_{W-N_{\rm p}-1}
\cdots
{\sigma}_{0}
{\sigma}_{N_{\rm s}-N_{\rm p}-1}
\cdots
{\sigma}_{W}]_2}^{
([
{\sigma}_{N_{\rm s}-1}
\cdots
{\sigma}_{N_{\rm s}-N_{\rm p}}
]_2)}.
\end{array}
\end{eqnarray}
{Finally, the following permutation of the components of the wavefunctions
completes the $W$-bit permutation of the distributed wavefunction:}
\begin{eqnarray}
\begin{array}{ll}
&v_{
[
{\sigma}_{W-N_{\rm p}-1}
\cdots
{\sigma}_{0}
{\sigma}_{N_{\rm s}-1}
\cdots
{\sigma}_{N_{\rm s}-N_{\rm p}}
{\sigma}_{N_{\rm s}-N_{\rm p}-1}
\cdots
{\sigma}_{W}]_2}^{
([
{\sigma}_{W-1}
\cdots
{\sigma}_{W-N_{\rm p}}]_2)}
\\
&=
{u'}_{
[
{\sigma}_{N_{\rm s}-1}
\cdots
{\sigma}_{N_{\rm s}-N_{\rm p}}
{\sigma}_{W-N_{\rm p}-1}
\cdots
{\sigma}_{0}
{\sigma}_{N_{\rm s}-N_{\rm p}-1}
\cdots
{\sigma}_{W}]_2}^{
([
{\sigma}_{W-1}
\cdots
{\sigma}_{W-N_{\rm p}}]_2)}.
\end{array}
\end{eqnarray}

Details of the continuous access will be reported elsewhere \cite{boost}.
The CMA method is implemented in Standard mode
for $S=1/2$ spins without total $S_z$ conservation in the present version of $\HPhi$
\footnote{
Since the implementation of the CMA method is more complicated than that of
the conventional algorithm explained in the previous section (Sec. \ref{ParaHvprod}),
the algorithm has not been implemented for other systems yet.
}.


\section{Benchmark results and performance}
\label{sec:Peformance}
In this section, we show benchmark results of $\HPhi$.
First, we examine the accuracy of the 
finite-temperature calculations based 
on the TPQ algorithm \cite{Sugiura2012} by 
comparing the results with the exact ensemble averages calculated
by the full diagonalization.
For small system sizes, 
we can easily perform the 
same calculation on our own PCs. 
Second, we show the parallelization efficiency of $\HPhi$ 
for large system sizes on supercomputers.
We have carried out TPQ simulations of two typical models, 
namely, an 18-site Hubbard model on a square lattice 
and a 36-site Heisenberg model on a kagome lattice, 
with changing numbers of threads and CPU cores,
where the number of CPU cores equals the number of threads times
the number of MPI processes.
We also show performance of 
the 
CMA method.

\subsection{{Benchmark results of TPQ simulations}}

%
To examine the validity of the TPQ calculation, 
we compute the temperature dependence of 
the doublon density of the
Hubbard model ($U/t = 8$) for an 8-site 
cluster with the periodic boundary condition.
The shape of the cluster is
illustrated in {the inset of Fig. \ref{fig_fdvstpq}}.
The Hubbard model is defined by setting $\mu=0$, $V_{ij}=0$,
$t_{ij}=t$ for the nearest-neighbor pairs of sites, 
$\langle i,j \rangle$, and $t_{ij}=0$ for further neighbor pairs of sites.
An example of the input file for the 8-site Hubbard model 
used in this calculation is shown as follows:
\begin{verbatim}
a0W     = 2
a0L     = 2
a1W     = -2
a1L     = 2
model   = "Fermion Hubbard"
method  = "TPQ"
lattice = "square lattice"
t       = 1.0
U       = 8.0
\end{verbatim}
%
\begin{figure}[!tb]
    \centering
    \includegraphics[width=7.0cm]{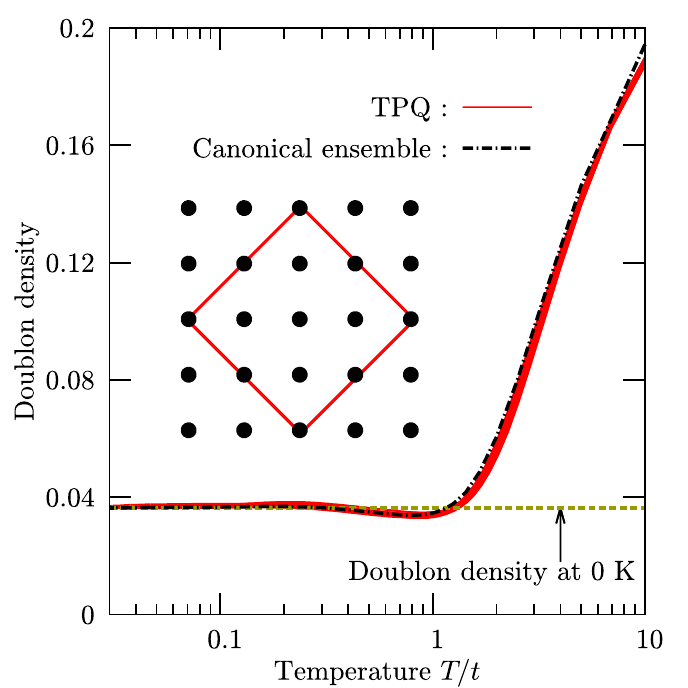}
    \caption{
      Temperature dependence of the doublon density calculated 
      by the TPQ state and the canonical ensemble obtained by the full diagonalization method
      on an 8-site Hubbard model.
      We perform 20 independent TPQ calculations 
      and depict all the results.
      The yellow horizontal line indicates the doublon density at zero temperature, 
      which is calculated by the Lanczos method.}
    \label{fig_fdvstpq}
\end{figure}
In Fig. \ref{fig_fdvstpq}, we show the 
temperature dependence of the doublon density,
$\langle {\hat n}_{\uparrow} {\hat n}_{\downarrow}\rangle$, calculated 
by the TPQ state and by the canonical ensemble.
We confirm that the TPQ calculations well reproduce
the results of the canonical ensemble average.
In addition to the finite-temperature
calculations, we  perform the ground state calculations by the
Lanczos method and the calculated 
doublon density at zero temperature
is plotted by the dashed line.
We also confirm that doublon density at low temperature
well converges to the value of the ground state.
All the results show the validity of the TPQ method.
%
%

\subsection{{Parallelization Efficiency}}

{Here, we carry out TPQ simulations for large 
system sizes with changing numbers of threads and processes
to examine parallelization efficiency of $\HPhi$.
We choose two typical models; a 
half-filled 18-site Hubbard model on a square lattice
and a 36-site Heisenberg model on a kagome lattice.
The speedup of the simulation 
for the 18-site Hubbard model with up to 3,072 cores
is measured on SGI ICE XA (Sekirei) at ISSP (Table \ref{tbl_spec}).
The speedup of the large-scale simulation 
for the 36-site Heisenberg model with up to 32,768 cores
is examined by using the K computer at RIKEN AICS (Table \ref{tbl_spec}).}

\begin{table*}[tb!]
  \begin{center}
    \caption{\label{tbl_spec} Specification of the single node on the supercomputer Sekirei
      \cite{sekirei}
      at ISSP and the K computer \cite{K-computer} at RIKEN AICS.
    }
    \begin{tabular}{ccc}
      \hline
      & Sekirei & K computer \\
      \hline
      CPU & Xeon E5-2680v3$\times$2 & SPARC 64\texttrademark VIIIfx \\
      Numer of cores per node & 24 & 8 \\
      Peak performance & 960 GFlops & 128 GFlops \\
      Main memory & 128 GB & 16 GB \\
      Memory band width & 136.4 GB/s &  64 GB/s \\
      Network topology & Enhanced hypercube & Six-dimensional mesh/torus\\
      Network band width & 7 GB/s $\times$ 2 & 5 GB/s $\times$ 2\\
      \hline
    \end{tabular}
  \end{center}
\end{table*}

\subsubsection{{18-site Hubbard model}}

\begin{figure}[!tb]
    \centering
    \includegraphics[width=7.0cm]{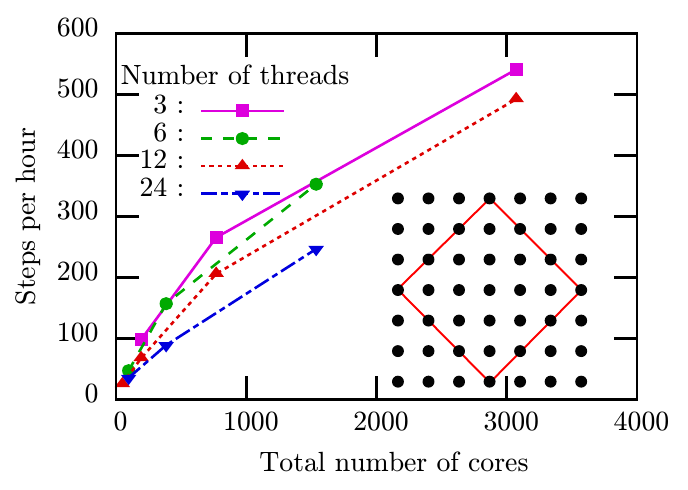}
    \caption{{Speedup of TPQ calculations with hybrid parallelization by using up to 3,072 cores on Sekirei.
    Here, we show TPQ steps per hour for the 18-site Hubbard cluster with finite $U/t$ ($U/t=8$) and total $S_z$ conservation ($\sum_{i=0}^{N_{\rm s}-1}S_{i}^z=0$)
    as functions of the total number of cores.
    We note that, due to changes in size of data transfer between MPI process, the thread number affects the TPQ steps per hour even when
    the total number of cores is fixed.
    The squares, circles, upward triangles, and downward triangles represent the results with 3 threads, 6 threads, 12 threads, and 24 threads, respectively.
    The inset shows the shape of the 18-site cluster used in this benchmark calculations.}}
    \label{fig_hubbard_speed}
\end{figure}

We perform the TPQ simulations for the half-filled 
18-site Hubbard model on the square lattice
illustrated in the inset of Fig. \ref{fig_hubbard_speed}.
Here, we employ the subspace of the 
Hilbert space that satisfies $\sum_{i=0}^{N_{\rm s}-1}S_{i}^z=0$.
Then, the dimension of the subspace is $(_{18}C_{9})^2=2,363,904,400$,
where $_{a}C_{b}$ represents the binomial coefficient.
The input file used in this calculations
is shown below:
\begin{verbatim}
a0W     = 3
a0L     = 3
a1W     = -3
a1L     = 3
model   = "Fermion Hubbard"
method  = "TPQ"
lattice = "square lattice"
t       = 1.0
U       = 8.0
nelec   = 18
2Sz     = 0
\end{verbatim}

In Fig. \ref{fig_hubbard_speed},
we can see significant acceleration caused by the increase of CPU cores.
This acceleration is almost linear up to 192 cores,
and is weakened as we further increase the number of cores.
This weakening of the acceleration comes from the load imbalance in the
$S^z$ conserved simulation.
For example, when we use 3 OpenMP threads and 1024 MPI processes,
the number of sites associated with the ranks of processes
( $N_p$ in Sec. \ref{sec_dist_wave} ) becomes 5, and
the number of sites in the subsystem in each process
( $N_{\rm Local}$ in Sec. \ref{sec_dist_wave}) becomes 13.
Each process has different numbers of up- and down- spin electrons.
In the process which has the largest Hilbert space,
there are six up-spin electrons and six down-spin electrons, and
the Hilbert space becomes $(_{13}C_{6})^2=2,944,656$.
On the other hand, the process which has the smallest Hilbert space
has nine up-spin electrons, nine down-spin electrons and
the $(_{13}C_{9})^2=511,225$ dimensional Hilbert space.
This large difference of the Hilbert-space dimension
between each process causes a load imbalance.
This effect becomes large as we increase the number of processes.

When we fix the total number of cores (for examples 1536 cores),
we can find that
the process-major computation (256 processes with 6 threads) is faster than
the thread-major computation (64 processes with 24 threads).
Since the numerical performance of both of them are equivalent,
this behavior seems strange at first sight.
The origin of difference 
might be explained as follows:
When we double the number of processes,
the amount of data communicated by each process at once
becomes a half of the original size,
and the communication time decreases.
On the other hand the communicated data-amount and
the communication time is unchanged even if we double the number of threads.
In other words, the algorithm described in Sec. \ref{ParaHvprod}
parallelizes calculation and also communication.
%

\subsubsection{36-site Heisenberg model}

\begin{figure}[!tb]
    \centering
    \includegraphics[width=7.0cm]{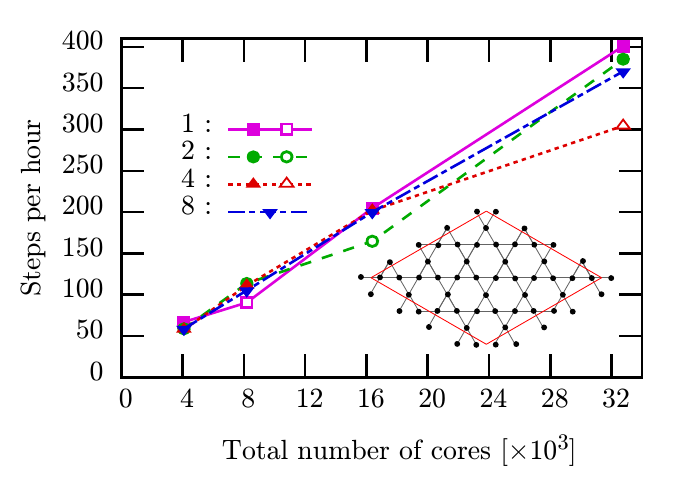}
    \caption{{Speedup of TPQ calculations with hybrid parallelization
        by using up to 32,768 cores on the K computer.
    Here, we show TPQ steps per hour for the 36-site Heisenberg cluster without total $S_z$ conservation as functions of the total number of cores.
    The squares, circles, upward triangles, and downward triangles represent the results with 1 thread, 2 threads, 4 threads, and 8 threads, respectively.
    The inset shows the shape of the 36-site cluster used in this benchmark calculations.}}
    \label{fig_kagome}
\end{figure}

We perform the TPQ calculations for
the 36-site Heisenberg model on the kagome lattice
with hybrid parallelization by using up to 32,768 cores.
Here, we employ the $2^{36}$-dimensional ($\sim 6\times 10^{10}$) Hilbert space
without the $S^z$ conservation.
Although the canonical ($S^z$ conserved) calculation is faster than the
grand-canonical ($S^z$ unconserved) calculation,
we sometimes employ the latter for the TPQ calculation.
The finite-size effect is much smaller in the grand-canonical TPQ state
than in the canonical TPQ state \cite{PhysRevB.90.121110}.
Figure \ref{fig_kagome} shows the speedup of this calculation on the K computer.
In contrast to the canonical calculation in the previous section,
we obtain the almost linear scaling even we use the very large number of cores
(more than ten thousands cores).
In the grand-canonical calculation, each process has a Hilbert space of the same size.
Therefore, there is no load imbalance
that appears prominently in the canonical calculation.
Due to the high throughput of the inter-node 
data transfer on the K computer, the parallelization 
efficiency from 4,096 cores to 32,768 cores reaches 82\%
and change in number of MPI processes does not largely affect
the speed.
Except for some cases (open symbols in Fig. \ref{fig_kagome}),
in the group of the same number of cores (for example, 16384 cores), 
we can not find any significant difference between the performance of
the thread-major computation (2048 processes with 8 threads) and
that of the process-major computation (16384 processes with 1 thread).
On the other hand when we calculate the same system by using
3,072 cores on Sekirei, 
the speed of
the computation by using 128 processes with 24 threads and
that speed by using 1024 processes with 3 threads become
29 steps / hour and 71 steps / hour, respectively;
we can find the advantage of the process-major computation on Sekirei
as in the previous section.
It is thought that
because the communication time on Sekirei is longer than
that on the K computer,
the parallelization of the communication
by increasing the number of processes is more effective
on Sekirei than on the K computer.

For some cases the speed becomes exceptionally slow
(corresponding conditions are shown as open symbols).
This slowdown occurs when the number of processes is 8,192,
and it remains even if we change the lattice to the one-dimensional chain.
We consider the reason of {the} slowdown as follows.
There are three effects on the computational speed by increasing processes:
first, the numerical cost per single process and 
the data size per single communication
decrease in inversely proportional to the number of processes (positive effect),
second, the frequency of communication increases (negative effect),
and,
finally, the load balancing becomes worse (negative effect).
Therefore we consider {the} slowdown comes from the competition of these effects.

\subsection{Benchmark of continuous-memory-access method}
In this section, we show benchmark results of
the continuous-memory-access parallelization algorithm, namely, the CMA method in $\HPhi$,
which is briefly explained in Sec. \ref{CA}.
The CMA method is particularly advantageous when the Hamiltonian does not possess
the U(1) symmetry or the SU(2) symmetry that make
the total charge and the total magnetization good quantum numbers.
A typical example is the Kitaev model discussed below. For such Hamiltonians,
we cannot use the canonical mode.
Especially, the CMA method is much faster 
than the conventional parallelization scheme 
{explained in Sec.\ref{ParaHvprod}}
when inter-site interactions are
dense and complicated.
This is because the cost for the computation and memory-transfer
for the bit permutation does not depend on the number of terms
or the complexity of the Hamiltonian as long as the range of interaction is not too long,
while, in the conventional parallelization scheme, the cost for the Hamiltonian operations increases as it has more terms.
To compare performance of the 
CMA method and the conventional parallelization scheme, here,
we carried out TPQ simulations of a spin Hamiltonian with
complicated spin-spin interactions.

We show the benchmark results of the TPQ simulation 
for a spin Hamiltonians of iridium oxide, Na$_2$IrO$_3$,
which was derived by utilizing {\it ab initio} 
electronic band structures and many-body perturbation theory \cite{Yamaji2014}.
The Hamiltonian is defined on a honeycomb structure and has 
complicated inter-site spin-spin interactions
that can be classified into
five parts as follows:
\begin{eqnarray}
\hat{\mathcal{H}}=\hat{\mathcal{H}}_{\rm K} + \hat{\mathcal{H}}_{\rm J} + \hat{\mathcal{H}}_{\rm off} + {\hat{\mathcal{H}}_{2}} + \hat{\mathcal{H}}_{3},
\label{ham_Na2IrO3}
\end{eqnarray}
where the first term $\hat{\mathcal{H}}_{\rm K}$ represents the Kitaev Hamiltonian \cite{AKitaev2006},
the second term $\hat{\mathcal{H}}_{\rm J}$ describes diagonal exchange couplings between nearest-neighbor (n.n.) spins,
and the third term $\hat{\mathcal{H}}_{\rm off}$ describes complicated off-diagonal spin-spin interactions.
The fourth term $\hat{\mathcal{H}}_2$ 
and the fifth term $\hat{\mathcal{H}}_3$ describe the second neighbor (2nd n.) interactions and
the third neighbor (3rd n.) interactions, respectively. 
Details of the Hamiltonians are given in the previous study \cite{Yamaji2014}.

\begin{figure}[bt!]
    \centering
    \includegraphics[width=7.0cm]{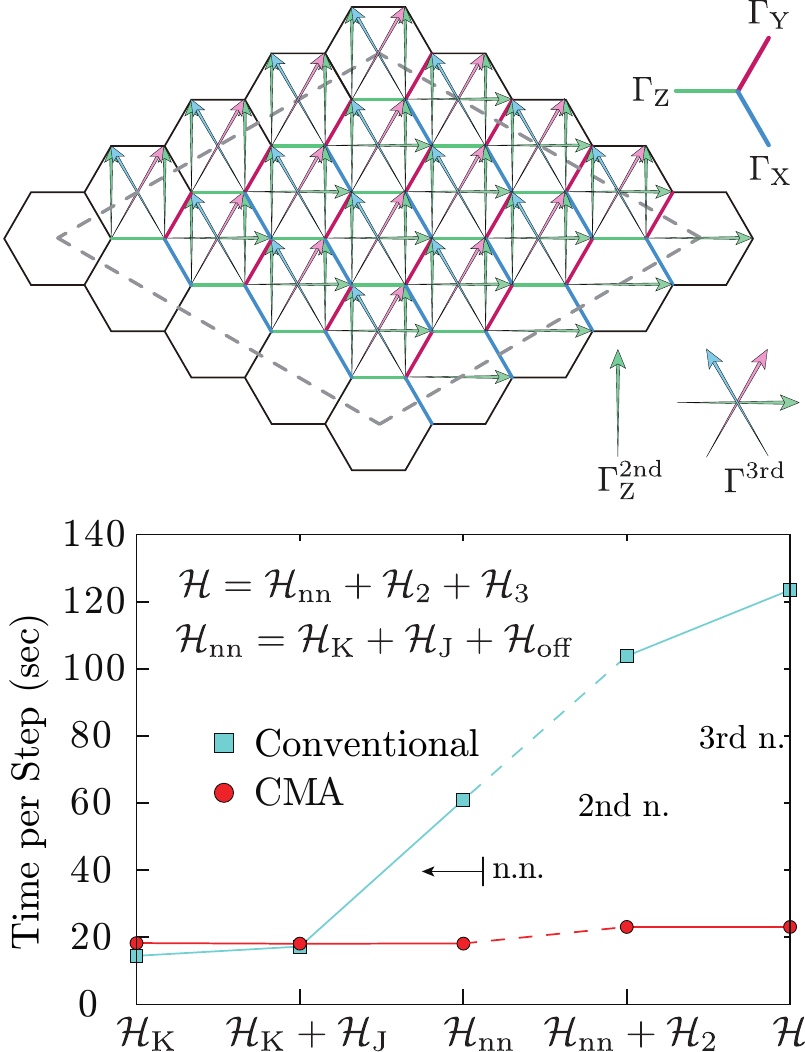}
    \caption{Lattice structure and bonds for $\hat{\mathcal{H}}$ (upper panel) 
      and elapsed time per TPQ step for the conventional butterfly
      and the CMA method (lower panel).
    The three different nearest-neighbor (n.n.) bonds, 
    $\Gamma_{\rm X}$, $\Gamma_{\rm Y}$, and $\Gamma_{\rm Z}$, 
    the second neighbor (2nd n.) bonds $\Gamma_{\rm Z}^{\rm 2nd}$,
    and the third neighbor (3rd n.) bonds $\Gamma^{\rm 3rd}$ are illustrated in the upper panel. 
    In the lower panel, the (cyan) squares denote the elapsed time per TPQ step 
    of the conventional scheme and the (red) circles denote
    the elapsed time per TPQ step of the CMA method.}
    \label{fig_Na2IrO3}
\end{figure}

In the lower panel of Fig. \ref{fig_Na2IrO3}, 
elapsed time per TPQ step for the conventional 
and the CMA method
is shown for a Kitaev Hamiltonian, $\hat{\mathcal{H}}_{\rm K}$, 
a Kitaev-Heisenberg Hamiltonian, $\hat{\mathcal{H}}_{\rm K}+\hat{\mathcal{H}}_{\rm J}$,
a n.n. Hamiltonian, $\hat{\mathcal{H}}_{\rm nn}=\hat{\mathcal{H}}_{\rm K}+\hat{\mathcal{H}}_{\rm J}+\hat{\mathcal{H}}_{\rm off}$,
a Hamiltonian including 2nd n. interactions, $\hat{\mathcal{H}}_{\rm nn}+\hat{\mathcal{H}}_{2}$, 
and the {\it ab initio} Hamiltonian $\hat{\mathcal{H}}=\hat{\mathcal{H}}_{\rm nn}+\hat{\mathcal{H}}_2+\hat{\mathcal{H}}_3$.
Here, we use 16 nodes (24 cores per node) 
on Sekirei at ISSP 
the TPQ simulation with 3 threads and 128 processes.
Even when the number of the coupling term increases, 
from the sparse Kitaev Hamiltonian $\hat{\mathcal{H}}_{\rm K}$ to
the dense and complicated Hamiltonian of Na$_2$IrO$_3$, $\hat{\mathcal{H}}$,
the elapsed time per TPQ step of the CMA method
does not show significant increase while 
the elapsed time of the conventional scheme shows significant increase.

Although, for the sparse Kitaev Hamiltonian $\hat{\mathcal{H}}_{\rm K}$, 
the conventional scheme shows slightly better 
performance than that of the CMA method,
the CMA method is three times faster 
than the conventional one even for the n.n. Hamiltonian $\hat{\mathcal{H}}_{\rm nn}$.
Furthermore, even though, from $\hat{\mathcal{H}}_{\rm nn}$ 
to $\hat{\mathcal{H}}_{\rm nn}+\hat{\mathcal{H}}_{2}$, the number of the bonds clearly increases, 
the elapsed time per TPQ step of the {CMA} method only shows slight increase.

%
%
\section{Summary}
To summarize, we explain the basic usage of
$\HPhi$ in Sec. 2. 
By preparing one file within ten lines,
for typical models in the condensed matter physics,
one can easily perform the exact diagonalization 
based on the Lanczos method
or finite-temperature calculations based on the
TPQ method.
By using
the same file, one can also perform 
the large-scale
calculations on modern supercomputers.
By 
properly editing the 
intermediate input files generated by Standard mode execution,
it is also easy to perform the similar calculations
for complicated Hamiltonians that describe the low-energy
physics of real materials.
This flexibility and user-friendly interface are the 
main advantage of $\HPhi$ and 
we expect that $\HPhi$ 
will be a useful tool
for broad spectrum of scientists 
including experimentalists and computer scientists.

In Sec. 3,
we explain the basic algorithms implemented in $\HPhi$. 
Although the Lanczos method and the TPQ method
are detailed in the literature, to make our paper self-contained,
we explain the essence of these methods.
Furthermore, in Sec. 4, we detail the parallelization methods 
of the multiplication of Hamiltonian to the wavefunctions;
one is the conventional butterfly method and another one is
the newly developed method.
We also show the benchmark results of $\HPhi$
on supercomputers in Sec. 5.
The results indicate that parallelization of
$\HPhi$ works well on these supercomputers 
and the 
computational time
is drastically reduced.

We also introduce the new algorithm for parallelization, i.e.,
the (CMA) method.
The main advantage of this method is that the 
speed does not depend on the number of terms in the Hamiltonians,
while the computational cost  
of the conventional butterfly method 
is proportional to the number of terms in the Hamiltonians.
Thus, the {CMA} method is suitable for treating
the complicated Hamiltonians for real materials.
Actually, for the low-energy model of Na$_{2}$IrO$_{3}$, 
we show that the speed of the {CMA} method
is about six times faster than that of the conventional method.
{So far,}
the CMA method is
applicable to the limited class of Hamiltonians and lattice geometries.
In addition, its efficiency 
largely depends on the models.
It is a remaining challenge 
to remove the limitation from the CMA method.

Recent development of the theoretical tools for obtaining
the low-energy models of real materials enable us 
directly compare the theoretical results and experimental results.
Although the limitation of the system size is severe,
the exact analyses based on the Lanczos method or
the TPQ method offer a firm basis for examining the validity 
of the theoretical results.
The present version of
$\HPhi$ can calculate 
only the static physical quantities, which
is difficult to be directly measured
in experiments.
Extending $\HPhi$ to calculate 
dynamical properties such as the dynamical spin/charge structure
factors and the optical conductivity 
is a promising way to make $\HPhi$ more useful.
Such an extension will be reported in the near future.

\section{Acknowledgements}

We would like to express our sincere gratitude to 
Prof. Hidetoshi Nishimori and Mr. Daisuke Tahara.
Implementation of the Lanczos algorithm in $\HPhi$ written in C is based on the
pioneering diagonalization package TITPACK ver. 2 written in Fortran by Prof. Nishimori.
For developing the user interface of $\HPhi$, we follow the design concept of
the user interface in the program for variational Monte Carlo method developed by Mr. Tahara.
A part of the user interface in $\HPhi$ is based on his original codes.
We would also like to thank the support from 
``{\it Project for advancement of software usability in materials science}" 
by The Institute for Solid State Physics, 
The University of Tokyo, for development of $\HPhi$ ver.1.0.
We thank the computational resources of the K computer 
provided by the RIKEN Advanced Institute for 
Computational Science through the General Trial Use project (hp160242).
This work was also supported by Grant-in-Aid for 
Scientific Research (15K17702, 16H06345, and 16K17746)
and Building of Consortia for the Development of Human Resources
in Science and Technology from the MEXT of Japan
and supported by PRESTO, JST.
We also thank numerical resources from the Supercomputer 
Center of the Institute for 
Solid State Physics, The University of Tokyo.

\appendix
\section{Details of implementation}
\subsection{Bogoliubov representation}\label{Bogo}
In the spin system,
the spin indices in input files of \verb|transfer|, \verb|InterAll|,
and correlation functions are specified by using the Bogoliubov representation.
Spin operators are written by using creation/annihilation operators as follows:
\begin{align}
  \hat{S}_{i}^{z} &= \sum_{\sigma = -S}^{S} \sigma \hat{c}_{i \sigma}^\dagger \hat{c}_{i \sigma}
  \\
  \hat{S}_{i}^+ &= \sum_{\sigma = -S}^{S-1} 
  \sqrt{S(S+1) - \sigma(\sigma+1)} 
  \hat{c}_{i \sigma+1}^\dagger \hat{c}_{i \sigma}
  \\
  \hat{S}_{i}^- &= \sum_{\sigma = -S}^{S-1} 
  \sqrt{S(S+1) - \sigma(\sigma+1)} 
  \hat{c}_{i \sigma}^\dagger \hat{c}_{i \sigma+1}
\end{align}

\subsection{Initial vector for the Lanczos method}
\label{ape:IniLz}
In the Lanczos method, an initial vector is specified 
with \verb|initial_iv|($\equiv r_s$) defined in an input file 
for Standard mode or a ModPara file for Expert mode. 
A type of an initial vector can be selected from 
a real number or complex number by using \verb|InitialVecType| in a ModPara file.
\begin{itemize}
\item{For canonical ensemble and \verb|initial_iv| $\geq 0$}

A non-zero component of a target of Hilbert space is given by
\begin{align}
(N_{\rm dim}/2 + r_s ) \% N_{\rm dim},
\end{align}
where $N_{\rm dim}$ is a total number of the Hilbert space 
and $N_{\rm dim}/2$ is added to avoid selecting the special configuration 
for a default value \verb|initial_iv| $=1$. 
When a type of an initial vector is selected as a real number, 
a coefficient value is given by $1$, 
while as a complex number, the value is given by $(1+i)/\sqrt{2}$.

\item{For grand canonical ensemble or \verb|initial_iv| $< 0$}

An initial vector is given by using a random generator, 
i.e. coefficients of all components for the initial vector 
is given by random numbers. The seed is calculated as 
\begin{align}
  123432+|r_s|+k_{\rm Thread}+N_{\rm Thread} \times k_{\rm Process},
  \label{fml_initialguess_lanczos}
\end{align}
where $r_s$ is a number given by an input file (parameter \verb|initial_iv|),
$k_{\rm Thread}$ is the thread ID,
$N_{\rm Thread}$ is the number of threads,
and $k_{\rm Process}$ is the process ID.
Therefore the initial vector depends both on \verb|initial_iv| and the number of processes.
Random numbers are generated by using SIMD-oriented Fast Mersenne Twister (dSFMT) \cite{Mutsuo2008}. 
\end{itemize}

\subsection{Convergence condition for Lanczos method}
\label{ape:ConvergenceLz}
In $\HPhi$,
we use \verb|dsyev| (routine of lapack)
for diagonalization of $T_{n}$.
We use the energy of the first excited state of $T_{n}$
as a criteria of convergence. 
In the standard setting,
after five Lanczos step,
we diagonalize $T_{n}$ every two Lanczos step.
If the energy of the first excited states agrees with
the previous energy within the required accuracy,
the Lanczos iteration finishes.
The accuracy of the convergence can be specified by 
\verb|LanczosEps| (ModPara file in Expert mode).

After obtaining the eigenvalues,
we again perform the Lanczos iteration
to obtain the eigenvector.
From the eigenvectors $|\Phi_n\rangle$,
we calculate 
energy $E_{n}=\langle \Phi_n|\Ham|\Phi_n\rangle $ and
variance $\Delta=\langle \Phi_n|\Ham^{2}|\Phi_n\rangle -(\langle \Phi_n|\Ham|\Phi_n\rangle)^2$.
If $E_{n}$ agrees with the eigenvalues obtained by the Lanczos iteration and 
$\Delta$ is smaller than the specified value,
we finish the diagonalization.

If the accuracy of Lanczos method is not enough,
we perform the CG method to obtain the eigenvector.
As an initial vector of the CG method,
we use the eigenvectors obtained by the Lanczos method in the standard setting.
This often accelerates the convergence.
\subsection{Initial vector for the TPQ method}
\label{ape:IniTPQ}
For TPQ method, an initial vector is given by using a random number generator,
i.e. coefficients of all components for the initial vector are given by random numbers. The seed is calculated as 
\begin{align}
123432+(n_{\rm run}+1)\times  |r_s|+k_{\rm Thread}+N_{\rm Thread} \times k_{\rm Process}
\end{align}
where $r_s$, $k_{\rm Thread}$, $N_{\rm Thread}$, and $k_{\rm Process}$ are the same
as those for the Lanczos method [Eqn. (\ref{fml_initialguess_lanczos})].
$n_{\rm run}$ indicates the counter of trials,
which is equal to or less than the total number of trials
\verb|NumAve| in an input file for Standard mode or a ModPara file for Expert mode.
We can select a type of initial vector from a real number or complex number by using \verb|InitialVecType| in a ModPara file.
$k_{\rm Thread}, N_{\rm Thread}, k_{\rm Process}$ indicate 
the thread ID, the number of threads, the process ID, respectively;
the initial vector depends both on \verb|initial_iv| and the number of processes.






{\bf References}

\begin{thebibliography}{10}
\expandafter\ifx\csname url\endcsname\relax
  \def\url#1{\texttt{#1}}\fi
\expandafter\ifx\csname urlprefix\endcsname\relax\def\urlprefix{URL }\fi
\expandafter\ifx\csname href\endcsname\relax
  \def\href#1#2{#2} \def\path#1{#1}\fi

\bibitem{PinesNozieres}
D.~Pines, P.~Nozi\`eres, The Theory of Quantum Liquids, Perseus Books
  Publishing, 1999.

\bibitem{Kittel}
C.~Kittel, Introduction to solid state physics, John Wiley \verb|&| Sons, Inc.,
  2005.

\bibitem{Dagotto}
E.~Dagotto, Rev. Mod. Phys. 66 (1994) 763--840.

\bibitem{titpack}
~http://www.stat.phys.titech.ac.jp/~nishimori/titpack2\_new/index-e.html.

\bibitem{kobepack}
~http://quattro.phys.sci.kobe-u.ac.jp/Kobe\_Pack/Kobe\_Pack.html.

\bibitem{spinpack}
~http://www-e.uni-magdeburg.de/jschulen/spin/.

\bibitem{1742-5468-2011-05-P05001}
B.~Bauer, L.~D. Carr, H.~G. Evertz, A.~Feiguin, J.~Freire, S.~Fuchs, L.~Gamper,
  J.~Gukelberger, E.~Gull, S.~Guertler, A.~Hehn, R.~Igarashi, S.~V. Isakov,
  D.~Koop, P.~N. Ma, P.~Mates, H.~Matsuo, O.~Parcollet, G.~PawﾅＰwski, J.~D.
  Picon, L.~Pollet, E.~Santos, V.~W. Scarola, U.~Schollwﾃｶck, C.~Silva,
  B.~Surer, S.~Todo, S.~Trebst, M.~Troyer, M.~L. Wall, P.~Werner, S.~Wessel,
  Journal of Statistical Mechanics: Theory and Experiment 2011~(05) (2011)
  P05001.
\newblock \href{http://stacks.iop.org/1742-5468/2011/i=05/a=P05001}{[link]}.
\newline\urlprefix\url{http://stacks.iop.org/1742-5468/2011/i=05/a=P05001}

\bibitem{Imada1986}
M.~Imada, M.~Takahashi, J. Phy. Soc. Jpn. 55~(10) (1986) 3354--3361.

\bibitem{FTLanczos}
J.~Jakli\ifmmode~\check{c}\else \v{c}\fi{}, P.~Prelov\ifmmode~\check{s}\else
  \v{s}\fi{}ek, Phys. Rev. B 49 (1994) 5065--5068.
\newblock \href {http://dx.doi.org/10.1103/PhysRevB.49.5065}
  {\path{doi:10.1103/PhysRevB.49.5065}},
  \href{http://link.aps.org/doi/10.1103/PhysRevB.49.5065}{[link]}.
\newline\urlprefix\url{http://link.aps.org/doi/10.1103/PhysRevB.49.5065}

\bibitem{Hams}
A.~Hams, H.~De~Raedt, Phys. Rev. E 62 (2000) 4365--4377.
\newblock \href {http://dx.doi.org/10.1103/PhysRevE.62.4365}
  {\path{doi:10.1103/PhysRevE.62.4365}},
  \href{http://link.aps.org/doi/10.1103/PhysRevE.62.4365}{[link]}.
\newline\urlprefix\url{http://link.aps.org/doi/10.1103/PhysRevE.62.4365}

\bibitem{Sugiura2012}
S.~Sugiura, A.~Shimizu, Phys. Rev. Lett. 108 (2012) 240401.

\bibitem{Yamaji2014}
Y.~Yamaji, Y.~Nomura, M.~Kurita, R.~Arita, M.~Imada, Phys. Rev. Lett. 113
  (2014) 107201.

\bibitem{pacheco1997parallel}
P.~Pacheco, Parallel Programming with MPI, Morgan Kaufmann Publishers, 1997.

\bibitem{MA}
~http://ma.cms-initiative.jp/en/application-list/hphi.

\bibitem{lapack}
E.~Anderson, Z.~Bai, C.~Bischof, L.~Blackford, J.~Demmel, J.~Dongarra,
  J.~Du~Croz, A.~Greenbaum, S.~Hammarling, A.~McKenney, D.~Sorensen, LAPACK
  Users' Guide, 3rd Edition, Society for Industrial and Applied Mathematics,
  1999.

\bibitem{MPI}
~http://www.mpi-forum.org/.

\bibitem{4052552}
K.~Martin, B.~Hoffman, IEEE Software 24~(1) (2007) 46--53.
\newblock \href {http://dx.doi.org/10.1109/MS.2007.5}
  {\path{doi:10.1109/MS.2007.5}}.

\bibitem{gnuplot}
~http://www.gnuplot.info/.

\bibitem{Lin}
H.~Q. Lin, Phys. Rev. B 42 (1990) 6561--6567.

\bibitem{hacker}
S.~H. Warren, Hacker's Delight (2nd Edition), Addison-Wesley, 2012.

\bibitem{SugiharaMurota_en}
M.~Sugihara, K.~Murota, Theoretical Numerical Linear Algebra, Iwanami Studies
  in Advanced Mathematics, Iwanami Shoten, Publishers, 2009.

\bibitem{Frommer}
A.~Frommer, Computing 70~(2) (2003) 87--109.

\bibitem{boost}
Y. Yamaji, unpublished.

\bibitem{sekirei}
~http://www.issp.u-tokyo.ac.jp/supercom/about-us/system/structure.

\bibitem{K-computer}
~http://www.aics.riken.jp/en/k-computer/system.

\bibitem{PhysRevB.90.121110}
M.~Hyuga, S.~Sugiura, K.~Sakai, A.~Shimizu, Phys. Rev. B 90 (2014) 121110.

\bibitem{AKitaev2006}
A.~Kitaev, Annals Phys. 321 (2006) 2--111.
\newblock \href {http://dx.doi.org/10.1016/j.aop.2005.10.005}
  {\path{doi:10.1016/j.aop.2005.10.005}}.

\bibitem{Mutsuo2008}
~http://www.math.sci.hiroshima-u.ac.jp/~m-mat/MT/SFMT.

\end{thebibliography}







\end{document}